\documentclass[10pt,journal,compsoc]{IEEEtran}

\ifCLASSOPTIONcompsoc
  \usepackage{color}
  \usepackage{cite}
  \usepackage{graphics} % for pdf, bitmapped graphics files
  \usepackage{epsfig} % for postscript graphics files
  \usepackage{mathptmx} % assumes new font selection scheme installed
  \usepackage{times} % assumes new font selection scheme installed
  \usepackage{amsmath} % assumes amsmath package installed
  \usepackage{amssymb}  % assumes amsmath package installed
  \usepackage{subfigure}
  \usepackage{diagbox}

\graphicspath{{eps_file/}}

\newtheorem{theorem}{\textbf{Theorem}}[section]
\newtheorem{myDef}{\textbf{Definition}}[section]
\newtheorem{lemma}{\textbf{Lemma}}[section]

\newtheorem{proposition}{\textbf{Proposition}}[section]

\else
\fi

\ifCLASSINFOpdf

\else

\fi

\begin{document}

\title{Resilient Consensus Against\\
Mobile Malicious Agents}

\author{Yuan~Wang,~\IEEEmembership{Member,~IEEE,}
        Hideaki~Ishii,~\IEEEmembership{Senior Member,~IEEE,}
        Fran\c{c}ois~Bonnet,
        and~Xavier~D\'{e}fago% <-this % stops a space
\IEEEcompsocitemizethanks{\IEEEcompsocthanksitem
Y.~Wang, H.~Ishii, F.~Bonnet and X.~D\'{e}fago are with the Department of Computer Science, Tokyo Institute of Technology, Tokyo/Yokohama, Japan.\protect\\
E-mail: wang@sc.dis.titech.ac.jp, \{ishii,bonnet,defago\}@c.titech.ac.jp}
\IEEEcompsocitemizethanks{\IEEEcompsocthanksitem
This work was supported in the part by the
JST CREST Grant No.~JPMJCR15K3 and by JSPS under Grant-in-Aid for
Scientific Research Grant No.~18H01460.}%
\thanks{Manuscript received March 2020.}
}

%\markboth{}{}

\IEEEtitleabstractindextext{%
\begin{abstract}
This paper addresses novel consensus problems in the presence of
adversaries that can move within the network and induce faulty
behaviors in the attacked agents. By adopting several mobile
adversary models from the computer science literature,
we develop protocols which can mitigate the influence of such
malicious agents. The algorithms follow the class of mean subsequence
reduced (MSR) algorithms, under which agents ignore the suspicious
values received from neighbors during their state updates.
Different from the static adversary models, even after the
adversaries move away, the infected agents may remain faulty
in their values, whose effects must be taken into account.
We develop conditions on the network structures for both the
complete and non-complete graph cases,
under which the proposed algorithms are guaranteed to attain
resilient consensus. Extensive simulations are carried out
over random graphs to verify the effectiveness of our approach
under uncertainties in the systems.
\end{abstract}

% Note that keywords are not normally used for peerreview papers.
}

% make the title area
\maketitle

\IEEEdisplaynontitleabstractindextext

\IEEEpeerreviewmaketitle

\section{Introduction}
\label{Section 1}

In recent years, together with the fast development of
communication networks, security problems have become
a critical issue in the domain of cyber-physical systems (CPSs).
In such systems, cyber attacks can cause damages not only
from having important information stolen,
but also from having physical equipments and devices manipulated,
which may lead to serious faults and dangerous accidents.
Security related problems have been investigated in
a wide range of disciplines including computer
science, control, signal processing, and robotics;
see, e.g., \cite{Chen2018,HeChen,Sandberg2015,robo2} and the references therein.
%\cite{Sandberg2015, Obenshain2016, Wheeler2019, Azevedo1998}.

In this paper, we follow the line of research on fault-tolerant
distributed algorithms \cite{Lamport,Lynch} and focus on resilient
consensus problems with real-valued states
(e.g., \cite{Bouzid2010,Dolev,Azadmanesh2000}).
Consensus problems form one of the most fundamental problems
in multi-agents systems \cite{Bullo,Mesbahi2010}.
There, agents locally communicate with neighbors for
arriving at the global objective to share a common value.
In uncertain environments, adversaries may attack
the agents to change their behaviors, which can result
in unexpected responses of the system and potentially
keep the non-faulty regular agents from reaching consensus
at a safe value.
Hence, it is of importance to guarantee that such regular
agents remain resilient and protect themselves from
adversarial attacks.

In particular, we deal with adversaries that can switch
the target agents from time to time. Such mobile adversaries
can cooperate in a worst-case manner by communicating
and collaborating with each other even if no direct link is present
among them in the network. On the other hand,
when the adversary leaves an attacked agent, it
may recover and become fault-free again. At the moment
of recovery, the value of such an agent may still be corrupted.
However, depending on the awareness of the agent itself,
it can take different actions. For example, it can use
only the neighbors' values for starting new in the consensus process.
Such a recovery may be performed by reboot or reset of
the system manually by the system operator or
automatically by devices such as watchdogs \cite{Ostrovsky1991}.

For mobile adversaries, several models have been proposed in
the literature \cite{Buhrman1995, Garay1994,Sasaki2013,Bonnet2016,Banu2011}.
These models are different in terms of
the timings of attacks for the adversaries and
the capabilities of the agents recovering from attacks or infections.
Recently, by \cite{Bonomi2019,Tseng2017, Sakavalas2018},
these studies have been extended to the case
where the agents' states take real values.
However, we must note that all of these studies are limited
in two aspects: One is that the networks are assumed to take
complete graph forms; such networks are very dense and require
resources for communications.
The other is that the adversaries are assumed to be
Byzantine. Such adversaries are considered
to be the worst type as they can freely manipulate their states
and are capable to send different messages to their neighbors.

The contribution of this work is threefold:
First, we extend the mobile adversary model in the
real-valued states case to the so-called \textit{malicious
adversary} models. Malicious agents form a subclass
of Byzantine agents and are slightly weaker in that they can only broadcast data,
that is, they send the same data to all neighbors.
Second, we propose novel protocols for achieving resilient consensus under
three different mobile malicious models.
The protocols follow the resilient approach known as the mean
subsequence reduced (MSR) algorithms \cite{Kieckhafer1994}.
In updating their state values, the agents ignore
suspicious values sent by other agents.
Third, we consider networks in non-complete graphs
and characterize the necessary connectivity structures for
the proposed MSR-based protocols to guarantee resilient
consensus.

The considered problem setting is natural from the viewpoint
of applications such as wireless sensor networks, where
agents communicate with a limited number of neighboring
agents and use broadcast transmissions.
Moreover, our results have been motivated by the recent advances
made in resilient consensus problems initiated by
\cite{LeBlanc2013} and \cite{Vaidya2012}. There, for MSR algorithms,
tight characterizations on the network structures have
been made by introducing the notion of graph robustness.
This approach has been extended in \cite{Dibaji2015,Dibaji2017}
for agent systems having higher-order dynamics together
with time delays in communication among the agents.
The work \cite{Dibaji2018} considers the case with quantized state
values, exhibiting that randomized algorithms can enhance the
applicability of the algorithms under asynchronous communication.
Further related studies can be found in, e.g.,
\cite{Fiore,robo1,Senej,Usevitch,wang:tcns2019,wang:ecc2019,Yuan,Zhang2015}.

As mentioned above, our work follows the line of research
in computer science on fault-tolerant consensus
in the presence of mobile adversary agents.
These works deal with Byzantine adversaries and agents
taking discrete state values.
The early work by \cite{Buhrman1995}
has proposed a model where the malicious agents can move and
switch their identities; when they move away,
the recovering agents are cured from infections immediately and
can be treated as regular in the next time step.
Another work by \cite{Garay1994} discusses a more general
model where the cured agent can detect the infection at the time
of recovery. Recently, other mobile adversary
models and resilient algorithms have been proposed by
\cite{Bonnet2016} and \cite{Sasaki2013}, where detection
of infection by the cured agents is not possible.
We extend these models to agents whose states take real values
under the malicious adversary model. In this case, in fact,
the two models in \cite{Bonnet2016} and \cite{Sasaki2013}
coincide, and thus, we study three different classes of
mobile adversary models in this paper.

The conventional MSR-based algorithms for the static
adversaries in, e.g., \cite{LeBlanc2013, Dibaji2017, Dibaji2018,Vaidya2012}
cannot guarantee resilient consensus
when the adversary agents are mobile.
This is mainly because the recovering agents require
special attention. An interesting aspect
here is that regular agents should not always trust
their own state values in their memory since they may
have been corrupted if the agents just recovered from infection
and are unable to detect this fact.
To mitigate the influence of such faults in the system,
in our modified MSR algorithms, the regular and recovering agents
treat their own values the same as their neighbors' values.
Thus, if their own values appear suspicious, they will be removed
and not used.
Compared with the conventional MSR algorithms for static models,
the agents must remove more values, and thus the network is required
to possess more connectivities. Clearly, this is the price to be
paid when the adversaries become mobile and thus more adversarial.

Early treatments of mobile adversary models can be found
in \cite{Ostrovsky1991}, which is motivated by processes of
virus spreading.
As pointed out by \cite{Yung2015}, it is valuable to consider
mobile adversaries models where at each time period, the number
of faulty parties is limited by a known bound,
but at each time step, the faulty parties can change
their identities. Such features lead to the phenomenon where
at each time, new faulty parties are introduced.
Meanwhile, the newly recovered parties can rejoin
the normal computation and then reach dynamic equilibrium.
In \cite{Li2015}, the approximate Byzantine consensus problem
is studied in a dynamic network,
where the nodes are mobile and can move.

The three mobile malicious models have different levels of
adversarial effects on the system. Our results clearly exhibit
trade-offs in that the required connectivities in the networks
become more strict as we assume the adversaries to have more power.
In other words, for ensuring higher resilience in the system,
the networks must possess more dense structures.
Our MSR-based approach allows us to reinterpret the recovery
mechanisms for one of the models (the M2 model of \cite{Garay1994}),
leading us to a novel protocol with a more relaxed condition on the
networks. Specifically, we let the recovering agents to take longer
time before starting to respond as regular agents.

This paper is organized as follows. In Section~\ref{Section 2},
some basic notions are introduced and then the problem considered
in this paper is formulated.
We propose four resilient consensus protocols for three
different mobile malicious models in
Sections~\ref{Section 3} to~\ref{Section 6}.
In our analysis, we provide conditions
for resilient consensus and, in particular, in terms of the required
network structures for both complete and non-complete graphs.
An illustrative example is provided in
Section~\ref{Section 7} to check the effectiveness of proposed
algorithms under uncertain environments where the theoretical
assumptions may not hold.
We give concluding remarks in Section~\ref{Section 8}.
A preliminary version of this paper will appear as a conference
paper \cite{wang:ifac2020}.
The current paper contains all proofs of the theoretical results
with further developments and discussions.
Extensive simulation studies are carried out as well.

\section{Problem Formulation}
\label{Section 2}

\subsection{General Notions}

Some basic notions on graphs are introduced for
the analysis that follows.
Denote by $\mathcal{G}=(\mathcal{V},\mathcal{E})$
the directed graph consisting of $n$ nodes, where the set of
nodes is $\mathcal{V}=\{1,2,\ldots,n\}$ and the set of
edges is $\mathcal{E} \subseteq \mathcal{V} \times \mathcal{V}$. The edge $(j,i) \in \mathcal{E}$ indicates that node~$j$ can send a message to node~$i$
and is called an incoming edge of node~$i$.
Let $\mathcal{N}_{i}=\{j\in\mathcal{V}:\,(j,i)\in \mathcal{E}\}$
be the set of (incoming) neighbors of node~$i$.
The degree $d_i$ of node~$i$ is the cardinality of its
neighbors set $\mathcal{N}_{i}$.

The path from node $i_1$ to node $i_p$ is denoted
as the sequence $(i_1, i_2, \ldots, i_p)$,
where $(i_j,i_{j+1}) \in \mathcal{E}$ for $j=1, \ldots, p-1$.
The graph $\mathcal{G}$ is said to have a spanning
tree if there exists a node from which there
are paths to all other nodes in this graph.
Moreover, the graph is said to be complete
if for each pair of nodes $i,j\in\mathcal{V}$,
there are bidirectional edges connecting them;
denote such a graph by $\mathcal{K}_n$.

To establish resilient consensus results, an important
topological notion is that of robustness of graphs \cite{LeBlanc2013}.

\begin{myDef}[Robust graphs]
\label{robust_graph}\rm
Given $r,s\in\{0,1,\ldots, n-1\}$
the graph $\mathcal{G}=(\mathcal{V},\mathcal{E})$
is called \emph{$(r,s)$-robust},
if for any two nonempty disjoint subsets $\mathcal{V}_{1},\mathcal{V}_{2}\subseteq\mathcal{V}$, one of
the following conditions is satisfied:
%\begin{enumerate}
\[
  \text{1)}~\mathcal{X}^{r}_{\mathcal{V}_{1}}=\mathcal{V}_{1},~~~
  \text{2)}~\mathcal{X}^{r}_{\mathcal{V}_{2}}=\mathcal{V}_{2},~~~
  \text{3)}~\bigl|\mathcal{X}^{r}_{\mathcal{V}_{1}}\bigr|
            +\bigl|\mathcal{X}^{r}_{\mathcal{V}_{2}}\bigr|\geq s,
\]
%\end{enumerate}
where $\mathcal{X}^{r}_{\mathcal{V}_{i}}$ is the set of all nodes
in $\mathcal{V}_{i}$ with
at least $r$ neighbors outside $\mathcal{V}_{i}$ for
$i=1,2$. Graphs with $(r,1)$-robustness are
said to be \emph{$r$-robust} as well.
\end{myDef}

We summarize some basic properties of robust graphs \cite{LeBlanc2013}.
Here, the ceil function $\lceil y \rceil$ gives the smallest integer
greater than or equal to $y$.

\begin{lemma} \label{lemma1}
An ($r,s$)-robust graph $\mathcal{G}$ satisfies the following:
\begin{enumerate}
  \item The graph $\mathcal{G}$ is ($r',s'$)-robust, where
     $0 \le r' \le r$, $1 \le s' \le s$, and in particular, it is $r$-robust.
  \item The graph $\mathcal{G}$ has a directed spanning tree.
        Moreover, it is $1$-robust if and only if it has a directed spanning tree.
  \item It holds $r \le \lceil n/2 \rceil$. Furthermore, if $r = \lceil n/2 \rceil$, $\mathcal{G}$ is a complete graph.
  \item The degree $d_i$ for $i \in \mathcal{V}$ is lower bounded as $d_i \ge r+s-1$ if $s<r$ and $d_i \ge 2r-2$ if $s \ge r$.
\end{enumerate}

Moreover, a graph $\mathcal{G}$ is $(r,s)$-robust if it is $(r+s-1)$-robust.
\end{lemma}

The usefulness of this notion in the context of consensus can
be seen from item 2) in the lemma; it is a generalization
of graphs containing directed spanning trees, which is central
to consensus problems without adversaries \cite{Bullo,Mesbahi2010}.

\subsection{Mobile Malicious Agents and Resilient Consensus}

We consider a multi-agent system with $n$ agents
interacting over the directed graph
$\mathcal{G}=(\mathcal{V},\mathcal{E})$.
Each node~$i\in\mathcal{V}$ has a state $x_i(k)$,
which takes a real value.
The objective of consensus is that
starting from initial values $x_i(0)$,
all agents update their states iteratively by
communicating with their neighbors so as to
arrive at the same value
as $\lim_{k\rightarrow\infty} |x_i(k)-x_j(k)|=0$ for
$i,j\in\mathcal{V}$.

In this paper, we study multi-agent systems
situated in an uncertain and even hostile environment.
Some of the agents are faulty and/or adversarial. Such agents
do not execute the given algorithm properly and may even update
their states arbitrarily with the intension to disturb
the ongoing consensus process. We introduce a new class for
such faulty agents, which is called
the \textit{mobile malicious model}.
Informally, this class has the following three features:
\begin{enumerate}
\item
Adversarial agents may transmit their false
states to their neighbors through broadcasting, i.e.,
all neighbors of a malicious agent receive the same data from it.
\item
The identity of the malicious agents can switch over
time. That is, an attacker may turn a non-adversarial agent
into a malicious one at certain time instants.
\item A malicious agent may recover and become regular.
The agent is said to be in the \textit{cured} status at that
moment. This happens
when the attacker decides to switch to another non-adversarial
agent.
\end{enumerate}

This model is said to be mobile to indicate that the attacker
may switch between different agents in infecting them.
In this work, we treat the mobile agents deterministically
though the mobile behaviors share similarities with
the stochastic models studied
for spreading processes of infectious diseases (e.g., \cite{Nowzari2016}).

We provide more notations and notions for
the mobile models considered in this paper.
At each time $k$, the set $\mathcal{V}$ of
nodes is partitioned into two subsets:
The set $\mathcal{R}(k)$ of regular agents
and the set $\mathcal{A}(k)$ of adversarial agents.
In the static case, both sets $\mathcal{R}(k)$ and $\mathcal{A}(k)$
remain invariant over time.

The faulty and abnormal behaviors
of the adversarial agents are defined below.

\begin{myDef}[Adversarial agents]
Three classes of adversarial agents are given as follows:
\begin{enumerate}
  \item (\emph{Byzantine}):
An adversarial agent $i\in\mathcal{A}(k)$
is said to be Byzantine if it makes updates in
its value $x_i(k)$
arbitrarily and may send different values to
its neighbors each time a transmission is made.
  \item (\emph{Malicious}):
An adversarial agent $i\in\mathcal{A}(k)$
is said to be malicious if it makes updates in its value $x_i(k)$
arbitrarily and sends the same value to all of
its neighbors each time a transmission is made.
  \item (\emph{Omissive}):
  An adversarial agent is said to make omissive faults
  if it does not send any value to any of its
  neighbors at times when transmissions are to be made.
\end{enumerate}
\end{myDef}

In this work, we focus on the class of
malicious agents, and thus our results cannot be directly
applied to networks with Byzantine agents.
It is clear that Byzantine adversary agents
have more capability than malicious agents.
However, the notion of malicious agents is relevant to many
applications. For example, in wireless sensor networks,
each sensor node communicates by broadcasting its data,
and hence its neighbors receive the same state data.
Also, in groups of mobile robots, the robots may determine
their neighbors' positions through measurements
by on-board sensors \cite{Mesbahi2010}.

Compared to the classical Byzantine models, malicious models have
received more attention only recently; see, e.g., \cite{LeBlanc2013, Dibaji2015}.
Different from the static version of malicious models studied there,
mobile adversaries can exhibit more
variety in their behaviors. As we discuss later,
we will adopt three classes of such mobile adversary models
from the literature, where Byzantine-type agents have been
studied.

Under the mobile adversary model, the identity
of the adversaries may switch, but we limit their
influence by bounding the total number of them in the
network over time. More specifically, we assume
the knowledge of an upper bound on the total number of
such agents. This is called
the $f$-total model as defined below.

\begin{myDef}[$\mathbf{f}$-total] \label{def2}
The mobile adversarial set
$\mathcal{A}(k)$ follows the \emph{$f$-total} model
if $|\mathcal{A}(k)|\leq f$ for all $k$,
where $f \in \mathbb{N}$.
\end{myDef}

For the multi-agent system in the presence
of mobile adversary agents, we introduce
the notion of resilient consensus.
Denote the maximum and minimum values of the states of
regular agents by
\begin{equation}
\begin{split}
\overline x(k)
  &= \max \{ x_i(k): i \in \mathcal{R}(k)\},\\
\underline x(k)
  &= \min \{ x_i(k): i \in \mathcal{R}(k)\},
\end{split}
\label{eqn:xbar}
\end{equation}
respectively. These values are well defined as long as regular agents
are present in the network (i.e., $\mathcal{R}(k)\neq \emptyset$).
To achieve resilient consensus, these are the values
that should eventually become the same in our problem setting.
%Note that these values are chosen only among the regular
%agents in $\mathcal{R}(k)$ at time $k$.

\begin{myDef}[Resilient consensus]
\label{def:consensus}
If for any possible sets and behaviors of the mobile malicious agents
in $\mathcal{A}(k)$
and any initial state values of the regular agents,
the following conditions are satisfied,
then the multi-agent system is said to reach resilient consensus:
\begin{enumerate}
  \item Safety condition:
Set the interval $\mathcal{S}=[\underline x(0),\overline x(0)]\subset \mathbb{R}$
containing the initial states of all regular agents at initial time.
Then, it holds $x_i(k)\in \mathcal{S}$ for all $ i \in \mathcal{R}(k)$,
$k\in \mathbb{Z}_{+}$.
%\textcolor{red}{HI: Note that the notation for the interval was changed from
%$\mathcal{S}$ to $\mathcal{S}(k)$.}
%\textcolor{blue}{Wang: I'm a little confused about this new definition. I think $x_i(k)\in \mathcal{S}(k)$ always hold based on the definition of $\underline x(k)$, $\overline x(k)$. Maybe the condition is $x_i(k+1)\in \mathcal{S}(k)$. In the static case, this $\mathcal{S}$ is a given initial interval and safety condition is $x_i(k)\in \mathcal{S}, i \in \mathcal{R}$. }
  \item Consensus condition: %For all $i,j \in \mathcal{R} $,
The regular agents eventually take the same value as
$\lim_{k \to \infty}\overline{x}(k) - \underline{x}(k) = 0$.
\end{enumerate}
\end{myDef}

The objective of this paper is to develop distributed
algorithms for the regular agents in the system to
reach resilient consensus as defined above. This problem is an extension
of those studied in
\cite{LeBlanc2013, Dibaji2015, Azadmanesh2000},
which are limited to the static adversary models.

Under the mobile adversary model, the notion of resilient
consensus is slightly different from the static case.
Since the agents may become malicious at any time,
even if after accomplishing consensus, an agent taking the consensus
value may suddenly change its value.
In fact, not only the agents in the adversary status, but also those in
the recovering status need not be in consensus with other regular agents.
In the definition above, however,
the safety interval $\mathcal{S}$ remains invariant to time
and is determined by the regular agents at the start time $k=0$.
%the adversarial as well as recovering agents to take arbitrary
%values even outside the safety interval $\mathcal{S}(k)$.

To mitigate the influence of the adversaries, we develop
modified versions of the so-called \textit{mean subsequence reduced (MSR)}
algorithms. For the static malicious model, such algorithms are known
to be capable of realizing resilient consensus. The basic update
rule for regular agents remains the same as outlined below.

\begin{figure*}[t]
    \centering
    \subfigure[Nodes]{%
    \label{fig6.00}
    \includegraphics[width=0.12\linewidth]{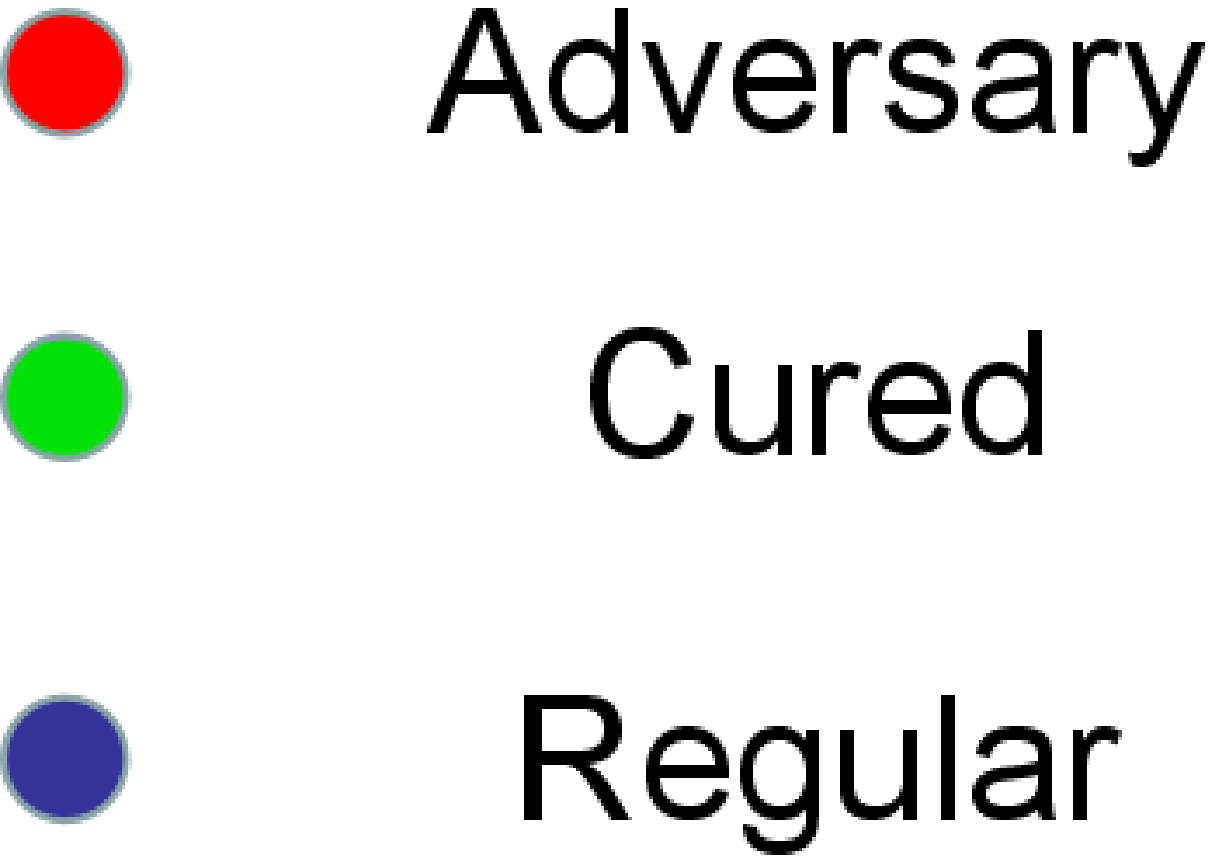}}
    \subfigure[Buhrman's mobile model (M1)]{%
    \label{fig6.01}
    \includegraphics[width=0.27\linewidth]{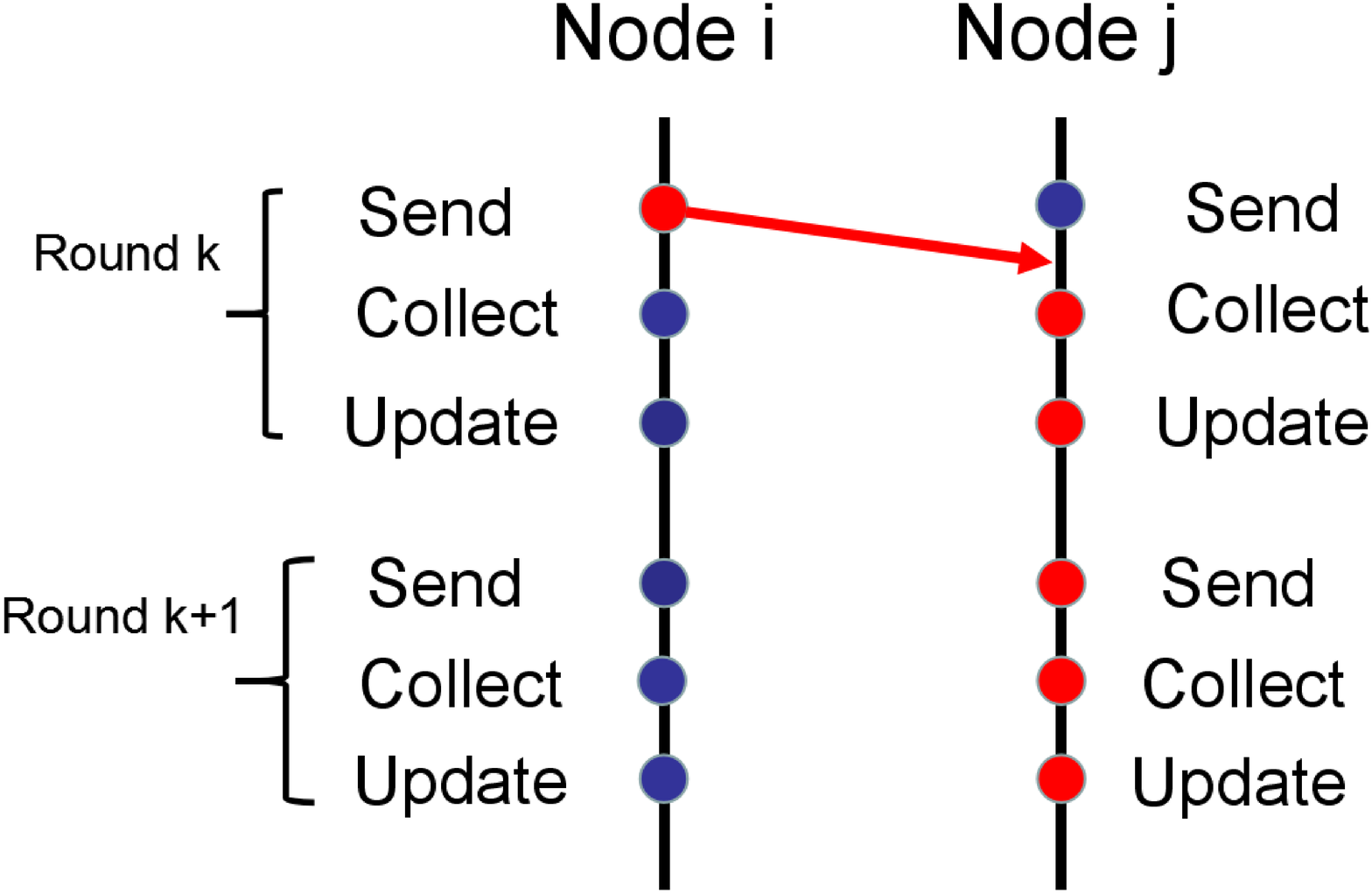}}
    \subfigure[Garay's mobile model (M2)]{%
    \label{fig6.02}\hspace*{2mm}
    \includegraphics[width=0.27\linewidth]{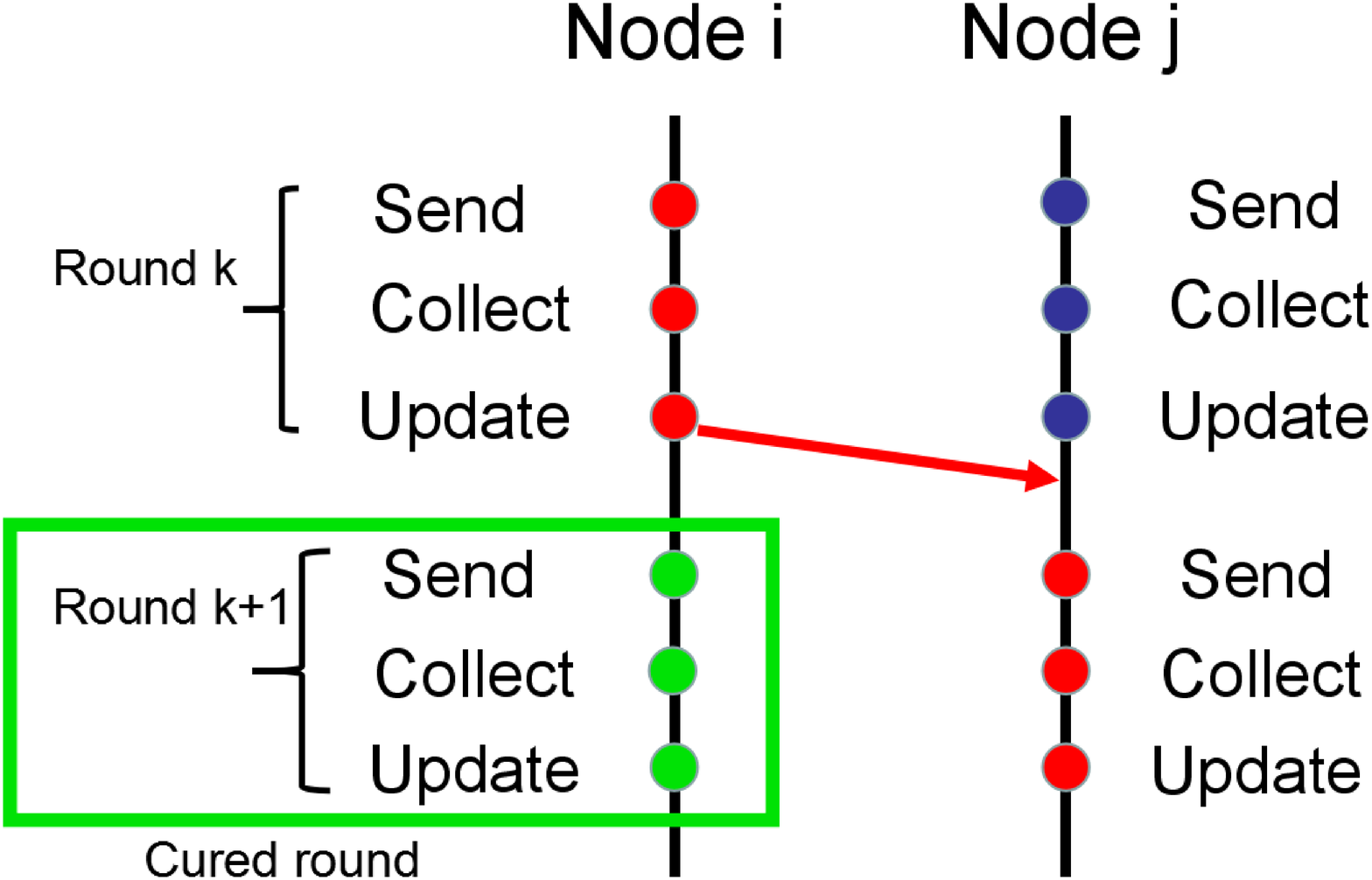}}
    \subfigure[Bonnet's mobile model (M3)]{%
    \label{fig6.03}\hspace*{2mm}
    \includegraphics[width=0.27\linewidth]{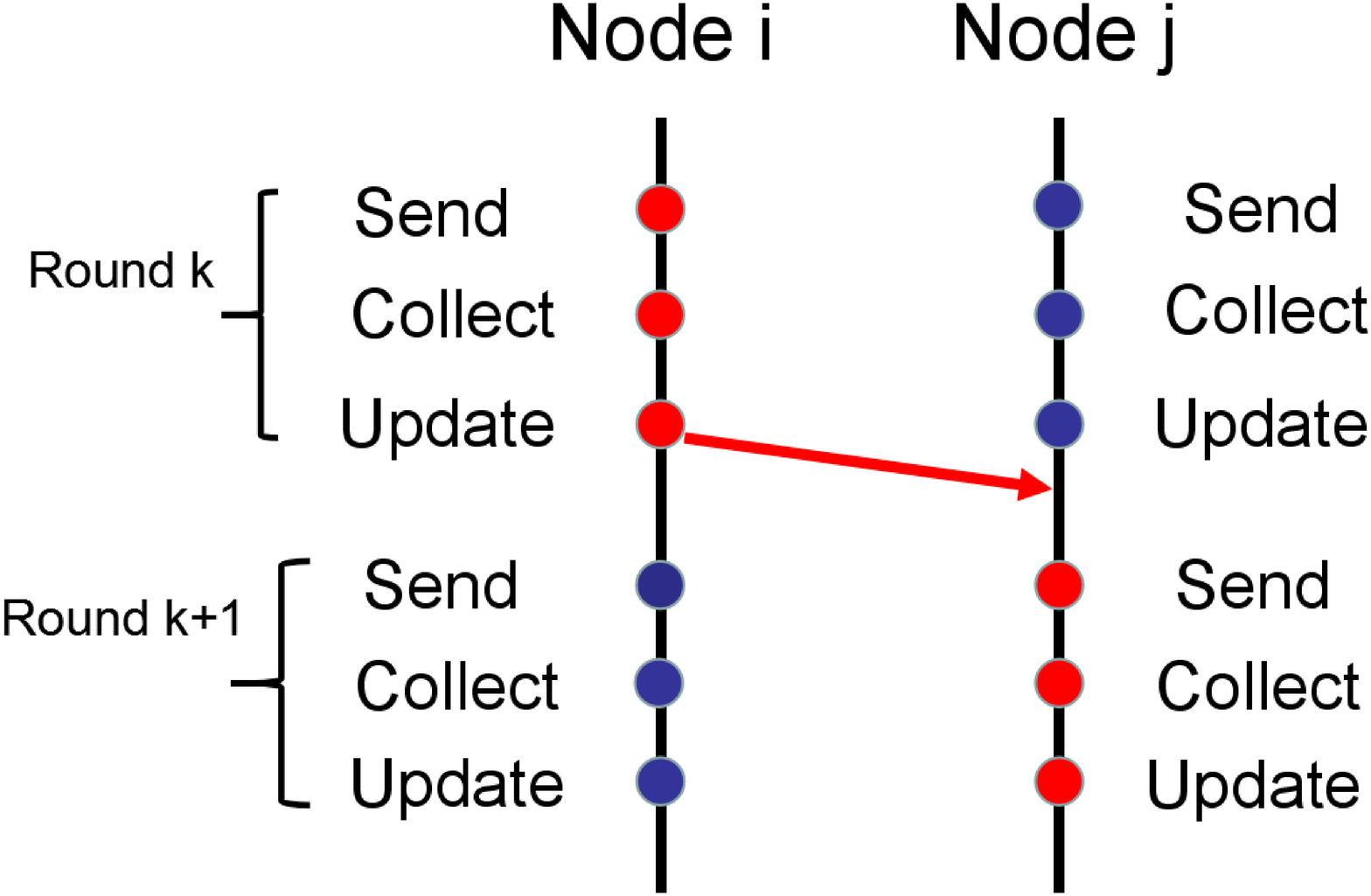}}
    \vspace*{-2mm}
    \caption{Mobile adversary models for the malicious agents case}
    \label{fig6}
\end{figure*}

For the state updates in the MSR algorithm, each agent executes
three basic steps \cite{LeBlanc2013}:
\textit{Send}, \textit{collect}, and \textit{update}.
At time (or \textit{round}) $k$, first,
a regular agent~$i$ broadcasts its current value $x_i(k)$
to its neighboring agents.
Second, it collects the values of the neighbor agents
$x_j(k)$ for $j\in \mathcal{N}_i$.
Third, after preprocessing to delete some of the neighbor
values, its value is updated to $x_i(k+1)$.
For the third step of state update, the update rule for
the state $x_i(k)$ of each regular agent~$i$ is given by
\begin{align}
  x_i(k+1)
   &= x_i(k)
     + \sum\limits_{j \in {\mathcal{M}_i}(k)} a_{ij}(k)
       \left(
         x_j(k) - x_i(k)
       \right),
\label{eqn:msr1}
\end{align}
where the weights must satisfy
$a_{ij}\in[\gamma,1)$ with $\gamma\in(0,1/2)$ and
$\sum_{j \in {\mathcal{M}_i}(k)} a_{ij}(k)\leq1$.
Here, $\mathcal{M}_i(k)$ denotes the subset of agent~$i$'s
neighbor set $\mathcal{N}_i$, whose states take safe values;
informally, among the neighbors, the $f$ largest and the
$f$ smallest values are removed to mitigate the influence
of the malicious agents.

It is known that to guarantee
resilient consensus by the MSR algorithm under the
$f$-total model, it is necessary
and sufficient that the network topology satisfies a
condition expressed in terms of its connectivity.
More specifically, the network must have the property
to be $(f+1,f+1)$-robust, as defined in Definition~\ref{robust_graph};
see, e.g., \cite{LeBlanc2013, Dibaji2015}.

However, we can show that mobile adversary agents
can easily destroy resilient consensus if the conventional
approach for the static $f$-total model is directly applied.
(For numerical simulations showing such properties,
see Section~\ref{sec:numerical:3}.)
One issue is related to the presence of the recovering nodes.
Suppose that, at one time, the adversary moves to
a different regular agent, which becomes malicious. At this moment,
the agent which was infected now recovers
and becomes regular. Such a recovering node might have a corrupted
value left in its memory from the attack. Note that in this round, there are
more than $f$ agents taking abnormal values in the network
even though each attacker is capable to infect only one agent at a time.

In our analysis, it is more convenient to use an alternative
expression of the update rule \eqref{eqn:msr1}.
Let the self-weight be given by
$ a_{ii}(k)
  = 1 - \sum_{j \in {\mathcal{M}_i}(k)} a_{ij}(k)$ and
the extended neighbor set
by $\mathcal{M}_i^+(k)\subset\{i\}\cup\mathcal{M}_i(k)$;
the set may contain the index of node~$i$ itself,
in which case $a_{ii}(k)\geq \gamma$ holds.
Then, we can rewrite the update rule \eqref{eqn:msr1} as
\begin{align}
  x_i(k+1)
  &= a_{ii}x_i(k)
     + \sum\limits_{j \in {\mathcal{M}_i}(k)}
         a_{ij}(k) x_j(k)\notag\\
  &= \sum\limits_{j \in {\mathcal{M}_i^+}(k)}
         a_{ij}(k) x_j(k).
\label{eqn:msr}
\end{align}

\subsection{Models for Mobile Malicious Behaviors}

Here, we introduce three classes of mobile malicious behaviors
denoted as models M1, M2, and M3.
The differences are related to what happens when an adversary
moves to another agent and, especially, to whether
the recovering agent is aware that it was attacked and its
state data may be corrupted.
These classes are taken from the literature in computer science
for the Byzantine adversaries.
We introduce the versions adapted for the malicious adversaries
case.

The three classes of mobile malicious models
are defined as follows (see Fig.~\ref{fig6}):
\begin{enumerate}
\item[\textit{M1}] \textit{Buhrman's model} \cite{Buhrman1995}:
The adversary may move away from an attacked
agent~$i$ only at the sending step in each round $k$ (Fig.~\ref{fig6.01}).
At such a round, agent~$i$ broadcasts its corrupted state
$x_i(k)$ to its neighbors, but then becomes recovered immediately;
thus, agent~$i$ collects and updates its state as a regular node.
For this reason, agent~$i$ will be classified as
regular in this round~$k$, i.e., $i\in\mathcal{R}(k)$.
If the adversary moved from agent~$i$ to another agent~$j$
after the send step, then we have $j\in\mathcal{A}(k)$.
It is important to note that at each round, there are
at most $f$ faulty values in the network.
\item[\textit{M2}] \textit{Garay's model} \cite{Garay1994}:
This model is characteristic in that each agent has an additional variable,
the \textit{cured flag} $\theta_i(k)$; initially, it is set
as $\theta_i(0)=0$.
The adversary can move away from an attacked agent~$i$ to
agent~$j$ at any step in each round $k$ (Fig.~\ref{fig6.02}).
Then, agent~$i$ is classified as adversarial at
round $k$, i.e., $i\in\mathcal{A}(k)$, and
as regular in the next round~$k+1$, i.e., $i\in\mathcal{R}(k+1)$.
In round~$k+1$, agent~$i$ is aware that it was infected and sets
its flag as $\theta_i(k+1)=1$.
It is set back to $\theta_i(k+1)=0$ after the update step in
round $k+1$.
At each round, there are at most $f$ faulty values and $f$
missing values in the network.
The cured flag can be used, e.g., to decide whether to
make transmissions or not.
\item[\textit{M3}] \textit{Bonnet's model} \cite{Bonnet2016}:
As in M2 above, under this model, the adversary agent can move away
from an attacked agent~$i$
at any step during each round~$k$ (Fig.~\ref{fig6.03}).
Thus, we have $i\in\mathcal{A}(k)$ and $i\in\mathcal{R}(k+1)$.
At round~$k+1$, agent~$i$ is in the recovering state, but
is not aware that it was infected. It
hence makes the next update as usual.
In this case, there are at most $2f$ faulty values
in the network: $f$ of them are due to attacks
and the remaining $f$ from cured agents like agent~$i$.
\end{enumerate}

To deal with each of these models, we provide four protocols
in the following sections.

\section{Protocol 1 for the M1 Model}
\label{Section 3}

\subsection{Modified MSR Algorithm 1}

Here, we present the first protocol for
the mobile adversaries, which is a modified version
of the MSR algorithm from, e.g., \cite{LeBlanc2013, Dibaji2015}.
It will be shown that this protocol is effective
to deal with mobile malicious agents
under the model M1 from \cite{Buhrman1995}.

\smallskip
\noindent
\textit{\textbf{Protocol 1.}}
%\begin{protocol}
At each round $k$, regular agent~$i\in\mathcal{R}(k)$
executes the following three steps:
\begin{enumerate}
  \item[1.] \emph{(Send)} Agent $i$ broadcasts its current value $x_i(k)$.
  \item[2.] \emph{(Collect)}  Agent $i$ collects the values $x_j(k)$
of neighbors $j \in \mathcal{N}_i$.
  \item[3.] \emph{(Update)}
(a)~Agent $i$ sorts the received values and
its own state value in a descending order.

(b)~After sorting, agent~$i$ deletes the $f$ largest
and the $f$ smallest values.
The deleted data will not be used in the update.
The set of indices of agents whose values remained
is written as $\mathcal{M}_i^+(k) \subset \{i\}\cup\mathcal{N}_i$.

(c)~Finally, agent~$i$ updates its value
by \eqref{eqn:msr}.
\end{enumerate}
%\end{protocol}

A unique feature of this algorithm is that
agent~$i$ might not use its own value.
This is because in Step~3, $2f$ values are deleted regardless of the
value of agent~$i$. By contrast, in the conventional
algorithms for the static adversary models in \cite{LeBlanc2013,Dibaji2015},
the number of values to be removed depends on the current
value of agent~$i$. Specifically, if agent~$i$'s
value is among the largest $f$ (respectively, the smallest $f$),
then only those greater (respectively, smaller) than $x_i(k)$
are deleted.

\subsection{Protocol 1 for the M1 Model: Complete Graphs}
We establish that with Protocol 1, we can achieve resilient
consensus under the M1 model.
Here, we first present the result
for networks in the complete graph form. More general graphs
will be treated in the next subsection.

\begin{proposition}\label{prop01}
Consider the multi-agent system whose network $\mathcal{G}$ forms
a complete graph.
Suppose that the mobile malicious agents follow the
$f$-total and M1 model.
Then, the regular agents using Protocol~1 reach resilient
consensus if and only if
$n \ge 2f +1$.
\end{proposition}

Before providing the proof, we introduce a few notations that
will be commonly used in the proofs of several results in the paper.
Denote the maximum difference among the values
of the regular nodes at time $k$ by
\begin{equation}
  V(k) = \overline{x}(k) - \underline{x}(k).
\label{eqn:V}
\end{equation}
Next, given $k$ with $V(k)>0$,
let the sequence $\varepsilon(k')$ for $k'\geq k$
be given by
\begin{equation}
  \varepsilon(k'+1) = \gamma \varepsilon(k'),
\label{eqn:epsilon}
\end{equation}
where $\varepsilon(k) = V(k)/2>0$.
Since $\gamma\in(0,1/2)$,
it holds $0 \le \varepsilon(k'+1) \le \varepsilon(k')$
for $k' \ge k$.
Then, define the two sets $\overline{\mathcal{X}}(k, k')$ and
$\underline{\mathcal{X}}(k, k')$ for $k'\geq k$ by
\begin{align}
 \overline{\mathcal{X}}(k, k')
   &=\bigl\{
       j \in {\mathcal{V}}:\
           x_j(k') > {\overline{x}(k)} - \varepsilon(k')
      \bigr\},
     \label{eqn:overlineX} \\
 \underline{\mathcal{X}}(k, k')
   &= \bigl\{
       j \in {\mathcal{V}}:\
           x_j(k') < {\underline{x}(k)} + \varepsilon(k')
      \bigr\}.
     \label{eqn:underlineX}
\end{align}
Here, let $\overline{\mathcal{X}}$ be the shorthand notation for
$\overline{\mathcal{X}}(k, k)$, and $\underline{\mathcal{X}}$ for
$\underline{\mathcal{X}}(k, k)$.
Notice that these are always disjoint and
contain at least one regular agent due to
$\varepsilon(k)>0$.

\smallskip\noindent
\textit{Proof of Proposition~\ref{prop01}}:~
%\begin{myproof}
The necessity part is straightforward. In the
update step in Protocol~1, there are $2f$ values removed by
each agent. Thus, if $n\leq 2f$, then there will be no value
left for updating the states of any of the regular agents.

For the sufficient part,
we must establish resilient consensus.
%To this end, denote the maximum difference among the values
%of the regular nodes at time $k$ by $V(k)=\overline{x}(k) - \underline{x}(k)$.
According to Definition~\ref{def:consensus},
we should show that the MSR-based Protocol~1
satisfy the two properties, the safety condition and the
consensus condition.

\iffalse
\begin{enumerate}
  \item [P1] For each regular agent~$i\in\mathcal{R}(k)$,
the updated value $x_i(k+1)$ should be inside the range
$[\underline{x}(k), \overline{x}(k)]$
determined by the regular values at time $k$.
  %\item [P2] The difference between regular agents is strictly smaller than the last time whenever the difference is nonzero, that is, $\overline{x}(k+1) - \underline{x}(k+1) < \overline{x}(k) - \underline{x}(k)$.
  \item [P2] The values of the regular agents eventually
becomes the same as
%$\lim_{k \to \infty} \overline{x}(k) - \underline{x}(k)=0$.
$\lim_{k \to \infty} V(k)=0$.
\end{enumerate}
\fi

To prove the safety condition, we show that
for each regular agent~$i\in\mathcal{R}(k)$,
the updated value $x_i(k+1)$ lies inside the range
$[\underline{x}(k), \overline{x}(k)]$
determined by the regular values at round $k$.
By definition of $\underline{x}(k)$ and $\overline{x}(k)$
in \eqref{eqn:xbar}, %at time $k$,
it holds
$x_i(k) \in [\underline{x}(k), \overline{x}(k)]$
for all regular agents $i\in\mathcal{R}(k)$.
Based on the deleting in Step~3 during the update in Protocol~1
and the $f$-total model,
for each regular~$i$,
if any neighbor $j\in\mathcal{N}_i$ has a value
$x_j(k)$ not in $[\underline{x}(k), \overline{x}(k)]$,
then this value is deleted as $j\notin\mathcal{M}_i^+(k)$.
%by agent~$i$ during the update step.
Similarly, if agent~$i$ is recovering, then
there is a chance that its own value $x_i(k)$ is corrupted
and is outside the interval
$[\underline{x}(k), \overline{x}(k)]$ due to attacks in
the previous round. However, in such cases,
this will be deleted in Protocol~1 as well.
Thus, by \eqref{eqn:msr},
the new state $x_i(k+1)$ is a convex combination of
values in $[\underline{x}(k), \overline{x}(k)]$ and thus
will remain in this interval.
%Therefore, we obtain
%$x_i(k+1) \in [\underline{x}(k), \overline{x}(k)]$.
%Consequently, the cured agents in $\mathcal{C}(k)$
%take regular values at the end of round $k+1$.

Next, we must show the consensus condition.
In this part, we fix $k\geq 0$.
Assume that consensus is not attained at this round $k$,
%i.e., $\overline{x}(k)-\underline{x}(k)>0$,
i.e., $V(k)>0$. Then, introduce the two sets
$\overline{\mathcal{X}}(k, k')$ and $\underline{\mathcal{X}}(k, k')$
from \eqref{eqn:overlineX} and \eqref{eqn:underlineX}, respectively.
Since the graph is complete with $n \ge 2f+1$,
for any agent~$i$ following Protocol~1, it sorts at least $2f+1$
values consisting of those received from its neighbors and its own,
and then removes $2f$ of them. Hence, at least one value from
them must remain, so the extended neighbor set
$\mathcal{M}_i^+(k)$ is nonempty.
Now, we partition this neighbor set $\mathcal{M}_i^+(k)$ into two
sets %with respect to $\overline{\mathcal{X}}}$
${\mathcal{M}_i^+(k) \setminus \overline{\mathcal{X}}}(k, k)$ and
${\mathcal{M}_i^+(k) \cap \overline{\mathcal{X}}}(k, k)$.
At least, one of them must be nonempty.
Suppose that ${\mathcal{M}_i^+(k) \setminus \overline{\mathcal{X}}}(k, k)$
is nonempty.
%Then, from the update equation \eqref{eqn:msr},
%it holds that for every regular agent $i$,
Then, from \eqref{eqn:msr}, it holds
\begin{flalign}
{x_i}\left( {k+1} \right)
& = \sum\limits_{j \in {\mathcal{M}_i^+(k) \cap \overline{\mathcal{X}}}} {{a_{ij}}} \left( {k} \right) {{x_j}\left( {k} \right)} +
\sum\limits_{j \in {\mathcal{M}_i^+(k) \setminus \overline{\mathcal{X}}}} {{a_{ij}}} \left( {k} \right) {{x_j}\left( {k} \right)} \nonumber \\
& \le (1- \gamma) \overline x(k)
   + \gamma (\overline x(k) - \varepsilon(k)) %\nonumber \\
 = \overline x(k) -\gamma \varepsilon(k)\nonumber\\
&= \overline x(k) - \varepsilon(k+1). %\nonumber.
\label{eqn:bound_xi}
\end{flalign}
%Take $\varepsilon(k+1) = \gamma \varepsilon(k)$.
%Then, in this case, all regular agents move
This indicates that agent~$i$ moves
outside the set $\overline{\mathcal{X}}(k,k+1)$ at round $k+1$.
Similarly, if ${\mathcal{M}_i^+(k) \cap \overline{\mathcal{X}}}(k,k)$
is nonempty, then %we have that all regular agents move
agent~$i$ moves
outside $\underline{\mathcal{X}}(k,k+1)$ at round $k+1$.

The argument above holds for any regular agent~$i$.
Thus, at round $k+1$, at least one of the sets
$\overline{\mathcal{X}}(k,k+1)$ and $\underline{\mathcal{X}}(k,k+1)$
does not contain any regular agent.
Suppose that the set $\overline{\mathcal{X}}(k,k+1)\cap\mathcal{R}(k+1)$
is empty. Then, we have
\begin{align}
 V(k+1)
& = \overline{x}(k+1) - \underline{x}(k+1) %\nonumber \\
 \le {\overline{x}(k)} - \varepsilon(k+1) - \underline{x}(k+1)
\nonumber \\
&\le \overline{x}(k) - \gamma \varepsilon(k) - \underline{x}(k)
%\nonumber \\
 \le V(k) - \gamma \varepsilon(k) \nonumber \\
&= \Bigl(1- \frac{\gamma}{2}\Bigr)V(k). %\nonumber
\label{eqn:V1}
\end{align}
The same bound holds if the other set
$\underline{\mathcal{X}}(k,k+1)\cap\mathcal{R}(k+1)$ is empty as well.
By repeating this argument for $k=0,1,\ldots$, we conclude that
$V(k)\leq (1-\gamma/2)^k V(0) \to 0$ as $k \to \infty$.
Thus, the consensus condition is established.
This completes the proof.
$\hfill\square$

\smallskip
This proposition can be seen as an extension of a result given in
\cite{Bonomi2019}, which deals with
the Byzantine-type mobile adversary model.
The condition there is $n\geq 3f+1$. This implies
that fewer adversaries can be tolerated in the network
compared to the malicious-type adversary case with $n\geq 2f+1$
given in the proposition above.
This is intuitive since Byzantine adversaries are more powerful.
The proof technique in \cite{Bonomi2019} is to transform
the problem so that a general result in \cite{Kieckhafer1994}
for static adversaries can be applied.
For Proposition~\ref{prop01}, we have proved
using arguments similar to those in \cite{LeBlanc2013,Dibaji2015},
which are also for the static case. We however remark that
the advantage of this approach is that it can be
extended to non-complete graph cases as we discuss next.

\subsection{Protocol 1 for the M1 Model: Non-complete Graphs}
Next, we demonstrate the effectiveness of the proposed Protocol~1
for the non-complete graph case
and provide a sufficient condition on the graph structure
for achieving resilient consensus under the M1 model.

\begin{theorem}\label{theorem02}
Consider the multi-agent system under the network $\mathcal{G}$
where the mobile malicious agents follow the $f$-total
and M1 model.
Then, the regular agents using Protocol~1 reach resilient
consensus if the following conditions are satisfied:
\begin{enumerate}
  \item[C1] $n \ge 4f+4$.
  \item[C2] For every agent $i$, the number of neighbors satisfies
$|\mathcal{N}_i| \ge 2f + 1 + n/{2} $.
\end{enumerate}
\end{theorem}

Note that condition C1 in the theorem is necessary for
condition C2 to hold. This can be easily shown.
In the graph~$\mathcal{G}$, the neighbor set of any node~$i$
satisfies $|\mathcal{N}_i| \leq n-1$.
Then, with C2, it follows $2f+1+n/2\leq |\mathcal{N}_i|\leq n-1$.
From these inequalities, we obtain $4f+4\leq n$.
We observe that when applied to complete graphs,
this result exhibits some
conservatism. The bound in Theorem~\ref{theorem02}
is $n\geq 4f+4$ whereas
in Proposition~\ref{prop01}, it is $n\geq 2f+1$.
Hence, the non-complete graph result requires a smaller
number of malicious agents in the network.

Before providing the proof of the theorem,
we transform the condition~C2 on the graph structure
in the lemma below.

\begin{lemma} \label{lemma01}
Consider the multi-agent network where the mobile malicious agents
follow the $f$-total model.
Then, under $n \ge 4f+4$, the condition C2
in Theorem~\ref{theorem02} holds if and only if
the following condition is satisfied:
\begin{enumerate}
  \item[C2$'$] There exists an integer $g \in[2f+2, n/2]$
such that for any $g$-agent subgraph $\mathcal{G}'$ of $\mathcal{G}$,
each agent inside $\mathcal{G}'$ has at least $2f+1$ neighbors
from the agents in the subgraph.
\end{enumerate}
\end{lemma}

%\noindent
\textit{Proof}:~
We first show that the graph $\mathcal{G}$
%satisfying condition~C2 in Theorem~\ref{theorem02}
satisfying C2 in Theorem~\ref{theorem02}
fulfils C2$'$. It suffices to show this
using $g = \lfloor n/2 \rfloor$.
%since if C2$'$ holds for $g = \lfloor n/2 \rfloor$, then it does so
%for any $g' \in[2f+2, n/2]$.
By deleting any $\lceil n/{2} \rceil$ agents
from $\mathcal{G}$,
we obtain a subgraph $\mathcal{G}'$ with the remaining agents,
where the number of nodes is equal to $\lfloor n/2 \rfloor=g$.
Based on C2,
we know that every agent in this subgraph has
at least $2f+1$ neighbors within the subgraph $\mathcal{G}'$.
Thus, C2$'$ holds.

Conversely, we establish that
if $\mathcal{G}$ satisfies C2$'$, then C2 holds.
We prove this by contradiction.
Suppose that in $\mathcal{G}$, there exists
an agent $i$ whose neighbor set $\mathcal{N}_i$
has cardinality $|\mathcal{N}_i|\leq 2f+n/2$.
We then arbitrarily remove $\lfloor n/2\rfloor$ neighbors
of agent~$i$ from the graph.
If there are fewer than $\lfloor n/2\rfloor$ neighbors, then
we remove all of them.
The remaining agents form a subgraph of $\mathcal{G}$
consisting of $\lceil n/2\rceil$ nodes.
However, for agent~$i$, the number of neighbors
contained in this subgraph is no greater
than $2f$.
%This implies that there is no
%$g \in[2f+2, n/2]$ for which C2$'$ holds.
This implies that for $g=\lceil n/2\rceil$,
the property C2$'$ does not hold. Moreover,
if it does not hold for $g=\lceil n/2 \rceil$,
it cannot hold for any smaller value of $g$.
$\mbox{}\hfill\square$

\smallskip
\noindent
\textit{Proof of Theorem \ref{theorem02}}:
Here, we outline the proof since it
follows along the lines similar to those
in the proof of Proposition~\ref{prop01}.
From there, the safety condition is clear, and it
remains to show the consensus condition.

Assume that consensus is not reached yet at round $k$,
and hence $V(k)>0$. We introduce the two sets
$\overline{\mathcal{X}}(k, k')$ and $\underline{\mathcal{X}}(k, k')$
from \eqref{eqn:overlineX} and \eqref{eqn:underlineX}, respectively.
%
\iffalse
\begin{flalign*}
\overline{\mathcal{X}}(k, k')
&=\{j \in {\mathcal{V}}: x_j(k')
      > {\overline{x}(k)} - \varepsilon(k') \}, \\%\label{eq-14} \\
\underline{\mathcal{X}}(k, k')
&=\{j \in {\mathcal{V}}: x_j(k')
      < {\underline{x}(k)} + \varepsilon(k') \}, %\label{eq-15}
\end{flalign*}
where $k'= k,k+1$.
Set $\varepsilon(k)=V(k)/2>0$
and $\varepsilon(k+1)=\gamma\varepsilon(k)$.
Denote by $\overline{\mathcal{X}}$ and $\underline{\mathcal{X}}$,
respectively the shorthand notations
for $\overline{\mathcal{X}}(k, k)$ and
$\underline{\mathcal{X}}(k, k)$.
\fi
%
We use the shorthand notations
$\overline{\mathcal{X}}$ and $\underline{\mathcal{X}}$,
respectively, for $\overline{\mathcal{X}}(k, k)$ and
$\underline{\mathcal{X}}(k, k)$.
Then, under conditions C1 and C2$'$ (from Lemma~\ref{lemma01}),
there are two cases to be considered:
(i)~$|\mathcal{V} \setminus \overline{\mathcal{X}}| \ge g$ and
(ii)~$|\mathcal{V} \setminus \overline{\mathcal{X}}| < g$.
In the following, we treat the case~(i) and
discuss the behavior of agents in the set $\overline{\mathcal{X}}$.
The other case~(ii) can be handled similarly
by focusing on the agents in $\underline{\mathcal{X}}$;
this is because
$|\underline{\mathcal{X}}|\leq|\mathcal{V} \setminus \overline{\mathcal{X}}| < g$
and, thus, it follows that
$|\mathcal{V} \setminus \underline{\mathcal{X}}|
\ge n-g\ge\lceil n/2 \rceil\ge g$.

For the case (i) with
$|\mathcal{V} \setminus \overline{\mathcal{X}}| \ge g$,
we look at the state behaviors
of the regular agents in $\mathcal{R}(k)$ (including those
in the recovering status).
We divide such agents into those in the sets
$\overline{\mathcal{X}}$ and
$\mathcal{V}\setminus\overline{\mathcal{X}}$.
First, we consider agent~$i \in \overline{\mathcal{X}}$.
Take a subgraph with $g$ agents in $\mathcal{G}$,
where $g-1$ agents are from
$\mathcal{V} \setminus \overline{\mathcal{X}}$
and the remaining one is agent~$i$.
Then, from the condition~C2$'$,
we know that agent~$i$ receives values from at
least $2f+1$ neighbors in this subgraph.
%$\mathcal{V} \setminus \overline{\mathcal{X}}$.
Thus, at this round $k$, after the removal of
$2f$ agents taking large or small state values
at the update step in Protocol~1,
the set
$\mathcal{M}_i^+(k) \cap \left( \mathcal{V} \setminus
\overline{\mathcal{X}} \right)
=\mathcal{M}_i^+(k) \setminus \overline{\mathcal{X}} $
is nonempty.
Then, under the update rule \eqref{eqn:msr} of
Protocol~1, the value $x_i(k+1)$ can be upper bounded as
in \eqref{eqn:bound_xi}.

Next, we consider the regular
agent~$i \in \mathcal{V} \setminus \overline{\mathcal{X}}$.
Due to $|\mathcal{V} \setminus \overline{\mathcal{X}}|\geq g$,
by the condition C2$'$,
agent~$i$ inside the subgraph
$\mathcal{V} \setminus \overline{\mathcal{X}}$
has at least $2f+1$ neighbors from the subgraph.
This implies that at the update in Protocol~1,
after the removal of $2f$ values,
agent~$i$ has one or more values left, that is,
the set $\mathcal{M}_i^+(k) \setminus \overline{\mathcal{X}}$
is nonempty. Hence, we have that
for each agent~$i \in \mathcal{V} \setminus \overline{\mathcal{X}}$,
the inequality \eqref{eqn:bound_xi} holds.
Therefore, we have shown that \eqref{eqn:bound_xi}
holds for each regular agent~$i\in\mathcal{R}(k)$.
It is now guaranteed that all regular agents are outside
the set $\overline{\mathcal{X}}(k,k+1)$ at round $k+1$.

Similarly, for the case (ii) with
$|\mathcal{V} \setminus \overline{\mathcal{X}}| < g$,
all regular agents are outside the set
$\underline{\mathcal{X}}(k,k+1)$ at round $k+1$.
Therefore, at round $k+1$, at least one of the sets
$\overline{\mathcal{X}}(k,k+1)$ and
$\underline{\mathcal{X}}(k,k+1)$
contains no regular agent.
By establishing the bound on $V(k)$
as in \eqref{eqn:V1}, we have
$V(k)\rightarrow 0$ as $k\rightarrow\infty$. \space \space \space
$\hfill\square$

\section{Protocol 2 for the M2 Model}
\label{Section 4}

We proceed to present another protocol that is effective for the M2 model
from \cite{Garay1994}.
This model is different from M1 in
that the recovering agents do not send their values to neighbors
since they are aware of having been infected.
This behavior can be considered as $f$-total omissive.
Hence, in the worst case under the M2 model, at each round,
there can be $f$-total malicious agents
and, in addition, $f$-total agents with omissive faults.

\smallskip\noindent
\textit{\textbf{Protocol 2.}}
%\begin{protocol}
At each round $k$, regular agent~$i\in\mathcal{R}(k)$
executes the following three steps:
\begin{enumerate}
  \item[1.] \emph{(Send)}
%If agent $i$ is not in the recovering
%status with the cured flag $\theta_i(k)=0$,
%then it broadcasts its current value $x_i(k)$.
If agent $i$ is in the regular status with
the cured flag $\theta_i(k)=0$,
then it broadcasts its current value $x_i(k)$.
Otherwise, with $\theta_i(k)=1$, it is in recovering
status and no transmission is made.
  \item[2.] \emph{(Collect)}  Agent $i$ collects the values $x_j(k)$
of neighbors $j \in \mathcal{N}_i$.
  \item[3.] \emph{(Update)}
(a)~If the cured flag is $\theta_i(k)=0$,
then agent~$i$ sorts the received values and
its own state value in a descending order.
Otherwise (i.e., $\theta_i(k)=1$),
agent~$i$ is recovering and sorts only
the received values.

(b)~After sorting, agent~$i$ deletes the $f$ largest
and the $f$ smallest values.
The deleted data will not be used in the update.
The set of indices of agents whose values remained
is written as $\mathcal{M}_i^+(k) \subset \{i\}\cup\mathcal{N}_i$.

(c)~Finally, agent~$i$ updates its value
by \eqref{eqn:msr}.
\end{enumerate}
%\end{protocol}

Similarly to Proposition~\ref{prop01},
we have the following result for networks
in the complete graph forms.

\begin{proposition}\label{prop03}
Consider the multi-agent system whose network forms
a complete graph.
Suppose that the mobile malicious agents follow the
$f$-total and M2 model.
Then, the regular agents using Protocol~1 reach resilient
consensus
if and only if the graph
satisfies $n \ge 3f +1$.
\end{proposition}

In the M2 model, there may be up to $f$ cured agents
(with $\theta_i(k)=1$), which are
not allowed to send their values to neighbors.
Hence, each regular node may not receive data from
some of its neighbors. Among the data received, up to $f$ of them
may be faulty. Protocol~2 is effective for this model
since each regular agent deletes $2f$ neighbor values in Step~3.
In comparison with M1, to guarantee its resilience for M2,
$f$ more neighbors are needed for each agent.

This argument also holds for the result extended
to non-complete graphs as shown in the theorem below.

\begin{theorem}\label{theorem04}
Consider the multi-agent system under the network $\mathcal{G}$
where the mobile malicious agents follow the $f$-total
and M2 model.
Then, regular agents using Protocol~2 reach resilient
consensus if the following conditions are satisfied:
\begin{enumerate}
  \item[C1] $n \ge 6f+4$.
  \item[C2] For every agent $i$, the number of neighbors satisfies
$|\mathcal{N}_i| \ge 3f + 1 + n/2$.
\end{enumerate}
\end{theorem}

We discuss the differences between Protocol~1 under M1
and Protocol~2 under M2.
Generally, the graph condition for M2 is stricter
than that for M1 because the agents in the
cured status complicate the system behavior. Furthermore, the
adversary agents in M2 are more powerful since they can
move at any step during the update rounds
while in M1,
they switch only at the send steps.

The main feature of M2 is that once an adversary agent moves
away, the recovering agent immediately knows that it was infected
and then avoids sending its value to neighbors.
In practice, this feature may not be easy to attain as
it requires the implementation of fault detection algorithms.
To deal with such an issue, we discuss yet another mobile
adversary model M3 in the next section. In this case,
detection of cured agents is not needed.
We propose another protocol to solve
resilient consensus for M3.

\section{Protocol 3 for the M3 Model}
\label{Section 5}

We outline our resilient protocol for the M3 model from \cite{Bonnet2016}.
Mobile adversaries under this model have more advantages since
the recovering agents do not know about their infection.
Hence, they send their values during the cured round
though they can be corrupted.
In this respect, the recovering agents can be considered as
additional $f$-total malicious agents in the network. As a
consequence, at each round, the regular agents may receive
at most $2f$ corrupted values.

Protocol~3 given below copes with the additional malicious
data in the M3 model.
It is a slightly modified version of Protocol~1.
Specifically, in Step~3 at each
round, $4f$ values are removed while in Protocol~1,
this number is $2f$.
We show that this protocol is effective to
deal with the mobile malicious agents under M3.

\smallskip\noindent
\textit{\textbf{Protocol 3.}}
%\begin{protocol}
At each round $k$, regular agent $i\in\mathcal{R}(k)$
executes the following three steps:
\begin{enumerate}
  \item[1.] \emph{(Send)} Agent $i$ broadcasts its current value $x_i(k)$.
  \item[2.] \emph{(Collect)}  Agent $i$ collects the values $x_j(k)$ of neighbors
$j \in \mathcal{N}_i$.
\item[3.] \emph{(Update)} (a)~Agent $i$ sorts the received values and
  its own value in a descending order.

(b)~After the sorting, agent~$i$ deletes the $2f$ largest and
the $2f$ smallest values.
The deleted data will not be used in the update.
The set of indices of agents whose values remained
is written as
$\mathcal{M}_i^+(k) \subset \{i\}\cup\mathcal{N}_i$.

(c)~Finally, agent $i \in \mathcal{R}$ updates its value by \eqref{eqn:msr}.
\end{enumerate}
%\end{protocol}

Since more values are deleted in Protocol~3,
the regular agents need more neighbors compared
with the networks for Protocols~1 and~2.
By an analysis similar to those in the previous sections,
we have the following results. The first is concerned
with networks in the complete graph form.

\begin{proposition}\label{prop05}
Consider the multi-agent system whose network forms
a complete graph.
Suppose that the mobile malicious agents follow the
$f$-total and M3 model.
Then, the regular agents using Protocol~2 reach resilient
consensus if and only if
$n \ge 4f +1$.
\end{proposition}

We can further deal with the non-complete
graph case as shown in the theorem below.

\renewcommand{\arraystretch}{1.05}
\begin{table}[t]
\centering
\fontsize{8}{10}\selectfont
\caption{Mobile adversary models and networks for resilient consensus}
\vspace*{-3mm}
\begin{tabular}{c|cccc}
\hline
%       & \multicolumn{2}{|c|}{Event-Triggered}
%       & Time- \\ \cline{2-3}
      &  Timing    & Awareness & \multicolumn{2}{c}{Network condition}\\
\cline{4-5}
Model &  of        & of being  & Complete & Non-complete \\
        &  infection &  cured  & graphs    & graphs \\ \hline
M1  & Send & --  & $n \geq 2f+1$ & $n \geq 4f+4$ \\
M2  & Any  & Yes & $n \geq 3f+1$ & $n \geq 6f+4$  \\
M3  & Any  & --  & $n \geq 4f+1$ & $n \geq 8f+4$ \\\hline
\end{tabular}
\label{tab1}
%\vspace*{-2mm}
\end{table}

\begin{theorem}\label{theorem06}
Consider the multi-agent system under the network $\mathcal{G}$
where the mobile malicious agents follow the $f$-total
and M3 model.
Then, regular agents using Protocol~3 reach resilient
consensus if the following conditions are satisfied:
\begin{enumerate}
  \item[C1] $n \ge 8f+4$.
  \item[C2] For every agent $i$, the number of neighbors satisfies
            $|\mathcal{N}_i| \ge 4f + 1 + n/2$.
\end{enumerate}
\end{theorem}

%\smallskip
We highlight the differences of the M3 model
and related results from those for M1 and M2.
First, we discuss the relation with M1.
Both M1 and M3 models do not require the functionality
to detect agents in the cured status.
However, the M3 model is more powerful since
in M1, the adversary agents can move only at the send step,
while in M3, they have more flexibility and can move at any step.
This difference results in a more restrictive condition
on the network structure to guarantee resilient consensus.
We observe that each agent
needs $2f$ more neighbors in M3 than in M1.

Next, we compare the M3 model with M2.
In both M2 and M3 models, an adversary agent can move
to another agent at any step during the rounds.
The difference comes from the detection ability in the
regular agents, and the agents in M2 are more capable
in this respect.
In M2, if a regular agent is infected by an adversary,
it becomes aware as soon as the adversary moves away.
In contrast, the regular
agents in M3 will never be aware of the infection,
and thus their response actions are limited.

As discussed above, we can find that the graph conditions
are related to the adversaries' power and defenders' ability.
In Table~\ref{tab1},
we summarize the properties of the three models
and the network conditions for the proposed protocols
obtained so far. Note that the conditions for the non-complete
graphs are partial (as only C1 in the corresponding theorems
are shown).

We note that in M2, the cured agents are guaranteed to
become regular in one round. However, to guarantee this
feature, we have to design the update rules carefully and
such update rules may require a more conservative graph condition.
In the following section, we follow the idea
of extending the curing round in M2 to multiple
rounds. Then, it becomes possible to design another class of
algorithms to guarantee resilient consensus.
%under more relaxed graph conditions.
%In the following section, we follow this idea.
%We extend the curing to two rounds
%and call them cure rounds 1 and 2.
%and different update rules in these rounds.
%The cured agents in both cure rounds do not send their values.
%For this model, we propose another protocol to achieve resilient consensus.

\section{Protocol 2A for the M2 Model}
\label{Section 6}

One of our objectives in this work is to relax the conditions
on network structures required for resilient consensus protocols.
Towards this end, we extend our approach for the M2 model.

%We will see that in Protocol~2 discussed in
%Section~\ref{Section 4}, there is room to improve the behavior
%of the cured agents so that they wait for one round before
%adopting the regular update rule.

  \begin{figure}[t]
        \centering\includegraphics[width=0.6 \linewidth]{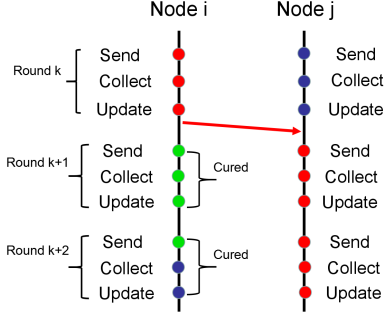}
    \vspace*{-2mm}
        \caption{Protocol 2A under the M2 model}
        \label{fig6.04}
%    \vspace*{-2mm}
  \end{figure}

\subsection{The M2 Model Revisited}

We introduce some modifications to the M2 model
and then develop a novel resilient consensus algorithm,
referred to as Protocol~2A.
Over Protocol~2 for the same mobile malicious model,
it has an advantage with respect to the requirement for
network connectivity.
In M2, the regular agents have the ability to detect whether they
were infected in the previous round; if so,
their values may still be infected and thus
must be ignored in the next update.
Recall the cured flag $\theta_i(k)$ introduced in the M2 model.
This variable is not part of the adversary model,
but rather a convention for the recovering agents.
Here, our approach is to generalize its function
by letting it take three values as $\theta_i(k)\in\{0,1,2\}$.
It is initially set as $\theta_i(0)=0$ for all $i$.

In the new algorithm here, when the adversary leaves agent~$i$
at round $k$, the cured agent~$i$ will take different actions in
the following two rounds $k+1$ and $k+2$ (Fig.~\ref{fig6.04}):
(i)~In the first round $k+1$, the flag is set as $\theta_i(k+1)=1$.
Agent~$i$ does
not send its value to the neighbors,
but only makes an update
by a rule different from that of regular agents.
(ii)~In the second round $k+2$, the flag becomes $\theta_i(k+1)=2$.
Agent~$i$ again does not send it value,
but applies the same update rule as the regular agents.
(iii)~After the update step in round $k+2$,
the value of agent~$i$ becomes regular.
The flag is set back to zero as $\theta_i(k+3)=0$.

Technically speaking, the cured agents with $\theta_i(k)=1$
are neither adversarial nor regular while
as discussed in Section~2,
cured agents with $\theta_i(k)=2$ may be considered regular
in their update rules; thus, we include them in $\mathcal{R}(k)$.
Hence, under this setting,
the node set $\mathcal{V}$ is partitioned into three sets
$\mathcal{A}(k)$, $\mathcal{C}(k)$, and $\mathcal{R}(k)$
at each round $k$, where
the cured agent set $\mathcal{C}(k)$ is given by
\begin{equation}
  \mathcal{C}(k)
   = \{i\in\mathcal{V}:\ \theta_i(k) = 1\}.
\label{eqn:Ck}
\end{equation}

%\textcolor{red}{HI: I think this discussion is needed
%for the definition of the safety interval.}
%\textcolor{blue}{Wang: In the new definition of the safety interval, the agents in cured round 2 is included, I think it is correct. Agents in cure round 2 follow the same update rule as regular agents and the upper and lower bounds $\overline x(k), \underline x(k)$ also include the agents in cured round 2. So we have $x_i(k+1)\in \mathcal{S}(k)$. Now let's avoid saying agents in cure round 2. It looks like, there are some special regular agents, that update normally but do not send their updated values.}

The details of Protocol~2A are described below.

\smallskip\noindent
\textit{\textbf{Protocol 2A.}}
At each round $k$, regular or cured
agent~$i\in\mathcal{R}(k)\cup\mathcal{C}(k)$
executes the following three steps:
\begin{enumerate}
  \item[1.] \emph{(Send)}
%If agent $i$ is not in the recovering status with
%the cured flag $\theta_i(k)=0$,
%then it broadcasts its current value $x_i(k)$.
If agent $i$ is in the regular status with
the cured flag $\theta_i(k)=0$,
then it broadcasts its current value $x_i(k)$.
Otherwise, with $\theta_i(k)=1$ or $2$, it is in recovering
status and no transmission is made.
  \item[2.] \emph{(Collect)}  Agent $i$ collects the values $x_j(k)$
of neighbors $j \in \mathcal{N}_i$.
  \item[3.] \emph{(Update)}
(a)~Agent~$i$ sorts the received values
%and its own state value
in a descending order.

(b1)~If the cured flag is $\theta_i(k)=0$ or
$\theta_i(k)=2$, then by comparing with its own
value $x_i(k)$, agent~$i$ deletes the $f$
largest and the $f$ smallest values from its neighbors.
If the number of values larger (or smaller)
than $x_i(k)$ is less than $f$, then all
of them are deleted.

(b2)~Otherwise, with $\theta_i(k)=1$,
agent~$i$ deletes the $f$ largest
and the $f$ smallest values and will not use its own
value.
%\textcolor{red}{HI: I am not very sure now if the last sentence is true.}
%\textcolor{blue}{Wang: Yes, this sentence is true. The agents in cure round 1 should first remove its own value, then delete the $f$ largest and the $f$ smallest values from the neighbors. However, the agents in cure round 2 should use its own value.}

(c) The deleted data will not be used in the update.
The set of indices of agents whose values remained
is written as $\mathcal{M}_i^+(k) \subset \{i\}\cup\mathcal{N}_i$.

(d)~Finally, agent~$i$ updates its value
by \eqref{eqn:msr}.
\end{enumerate}

%On the other hand, for the regular agents or the cured agents in cure round 2, they apply the conventional MSR update rule. More specifically, the cured agents in cure round 1 apply the modified MSR update rule as in Protocol 1, where the difference is that the regular agents in Protocol 4 delete less connections. We will see that Protocol 4 may guarantee the resilient consensus with less connections in non-complete graphs. Then we introduce the following theorem.

Note that in Step~3\,(b1) of Protocol~2A,
the agents may remove less than $2f$ values among those
received from the neighbors, which follows the MSR
approach \cite{LeBlanc2013,Dibaji2015}.
In fact, we can establish a sufficient condition for the
network topology to attain resilient consensus based on
the notion of robust graphs introduced in Section~2.

\subsection{Resilient Consensus Result}

The main result for Protocol~2A is
presented in the following theorem.

\begin{theorem}\label{theorem09}
Consider the multi-agent system under the network $\mathcal{G}$
where the the mobile malicious agents follow
the $f$-total and M2 model. Then, the regular agents
using Protocol~2A reach resilient consensus
if the graph is $(4f+1,2f+1)$-robust.
%The safety interval is given by
%$\mathcal{S} = \left[ \underline{x}(0), \overline{x}(0) \right]$.
\end{theorem}

%\smallskip
\noindent
\textit{Proof}:
We first show the safety condition part.
According to the M2 model, at any round $k$,
there are at most $f$ adversary agents and $f$ cured agents.
Since the graph is $(4f+1,2f+1)$-robust, by definition,
each agent has more than $4f+1$ neighbors.
In the update rule of Protocol~2A,
for both regular agents and cured agents,
at most $2f$ values are deleted
and up to $2f$ may be missing from omissive faults
(from the agents in the two cured rounds);
thus, their neighbor set $\mathcal{M}_i(k)$ used
in the update rule \eqref{eqn:msr} is nonempty for any $k$.

More specifically,
for each regular agent $i \in \mathcal{R}(k)$
(including the cured case with $\theta_i(k)=2$),
based on the Step~3\,(b1), we know that if any neighbor~$j$
satisfies $x_j(k) \notin [\underline{x}(k), \overline{x}(k)]$,
then it will be deleted.
So based on \eqref{eqn:msr}, we have that
$x_i(k+1) \in [\underline{x}(k), \overline{x}(k)]$.
For each cured agent $i\in\mathcal{C}(k)$ in \eqref{eqn:Ck},
its own value $x_i(k)$ may be infected, and
%the possible corrupted value at round $k$
%also comes from the previous infected value $x_i(k)$.
it will be deleted in Step~3\,(b2); hence
in the update rule \eqref{eqn:msr}, we have $a_{ii}(k) = 0$.
%where $a_{ii}(0) \neq 0$ only when
%${x}_i(0) \in [\underline{x}(0), \overline{x}(0)]$.
Thus, it follows $x_i(k+1) \in [\underline{x}(k), \overline{x}(k)]$.
This implies that $\overline{x}(k)$ and $\underline{x}(k)$
are nonincreasing and nondecreasing functions of round $k$.
From the above, it follows that
the regular and cured agents remain within
the safety interval $\mathcal{S}=[\underline{x}(0), \overline{x}(0)]$
at the end of round $k$.

In the rest of the proof, we must prove the consensus condition.
%We first discuss the update of regular agents.
Here, we consider the system behavior for a fixed round $k$.
We use $V(k)$ in \eqref{eqn:V} and the two sets
$\overline{\mathcal{X}}(k,k')$ and
$\underline{\mathcal{X}}(k,k')$ given in \eqref{eqn:overlineX}
and \eqref{eqn:underlineX}, respectively,
where $k'\geq k$.
By definition, these two sets are disjoint and nonempty.
Thus, from the assumption of $(4f+1,2f+1)$-robustness,
we have the following three cases:
\begin{enumerate}
  \item All agents in $\overline {\mathcal{X}}(k,k')$ have at least $4f+1$
        neighbors from outside the set.
  \item All agents in $\underline {\mathcal{X}}(k,k')$ have at
        least $4f+1$ neighbors from outside the set.
  \item The total number of agents in $\overline {\mathcal{X}}(k,k')$ and
        $\underline {\mathcal{X}}(k,k')$ that have at least $4f+1$
        neighbors outside the set to which they belong is
        no smaller than $2f+1$.
\end{enumerate}

\begin{figure}[t]
\centering
\includegraphics[width=0.9\linewidth]{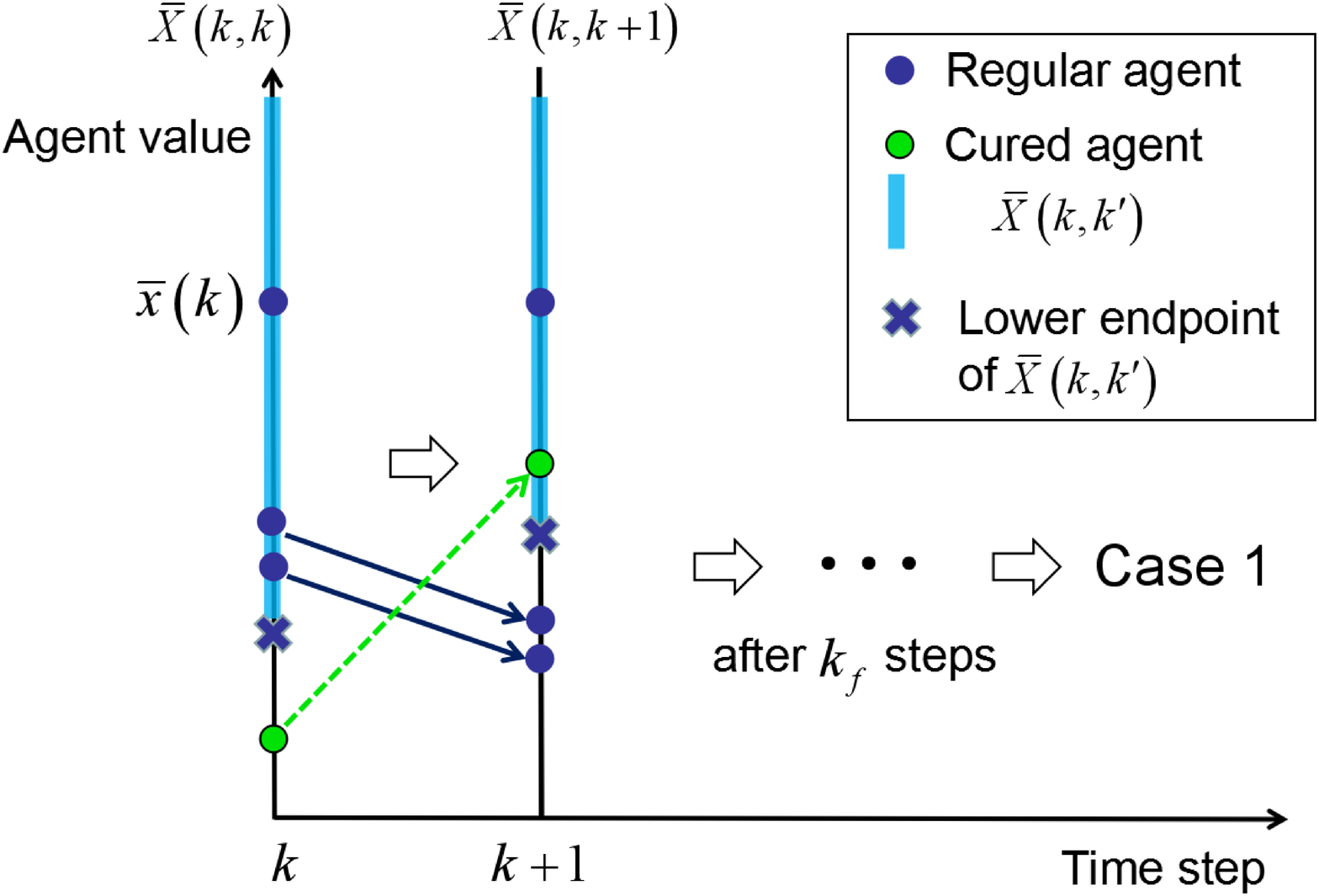}
\vspace*{-2mm}
\caption{Updates for regular and cured agents in case 3}
\label{fig05}
%\vspace*{-2mm}
\end{figure}

Here, we claim that case~3 above
will eventually reduce to case~1 or~2 in a future time;
this case is illustrated in Fig.~\ref{fig05}.
%Thus, we first discuss this case.
In particular, we show that the number
of regular and cured agents
in $\overline {\mathcal{X}}(k,k')$ at round $k'\geq k$
%$\overline {\mathcal{X}}(k,k + 1)$ at time $k+1$.
decreases over the rounds.
Note that under case~3, there are at least $f+1$ regular
agents (including cured agents with $\theta_i(k)=2$)
in total in $\overline {\mathcal{X}}(k,k)$
and $\underline {\mathcal{X}}(k,k)$
that have at least $4f+1$ neighbors from outside the
corresponding sets. We consider the regular agents
and the cured agents separately.

%Note that the agents in $\underline {\mathcal{X}}(k,k,\varepsilon(k))$ can be similarly analyzed.
%since ${\varepsilon}\left( {k + 1} \right) = \gamma {\varepsilon}\left( k \right)$,

First, we study the regular agents,
which follow the updates in Step 3\,(b1).
Take one agent~$i\in\mathcal{R}(k)\cap\overline {\mathcal{X}}(k,k)$
having at least $4f+1$ neighbors
from outside $\overline {\mathcal{X}}(k,k)$.
In the update rule \eqref{eqn:msr},
partition the agents in $\mathcal{M}_i(k)$ into two parts as
$\mathcal{M}_i(k)\cap \overline{\mathcal{X}}(k,k)$ and
$\mathcal{M}_i(k)\setminus \overline {\mathcal{X}}(k,k)$.
Then, we can write
\begin{align}\label{eq-09}
  x_i(k+1)
   &= a_{ii}(k)x_i(k)
       + \sum\limits_{j \in \mathcal{M}_i(k)\cap\overline{\mathcal{X}}}
          a_{ij}(k) x_j(k) \nonumber \\
   &{\kern 24mm} \mbox{}
       + \sum\limits_{j \in \mathcal{M}_i(k)\setminus \overline {\mathcal{X}}}
          a_{ij}(k) x_j(k),
\end{align}
where $\overline {\mathcal{X}}$ is the shorthand notation of
$\overline {\mathcal{X}}(k,k)$.
%Suppose that this agent~$i$ has at least $4f+1$ neighbors
%from outside $\overline {\mathcal{X}}(k,k)$.
Among the neighbors, there are at most $2f$ agents
with omissive faults,
and at most $f$ neighbors are removed.
%recall that these agents' values are no greater
%than $\overline{x}(k)-\varepsilon(k)$.
It is clear that the set
$\mathcal{M}_i(k) \setminus \overline {\mathcal{X}}(k,k)$
contains regular agents.
Thus, we have
\begin{align}\label{eq-10}
  x_i(k+1)
   &\le a_{ii}(k) \overline {x}(k)
        + \sum\limits_{j \in \mathcal{M}_i(k) \cap \overline {\mathcal{X}}}
         a_{ij}(k) \overline {x}(k) \nonumber \\
   & {\kern 24mm}\mbox{}
        + \sum\limits_{j \in \mathcal{M}_i(k) \setminus \overline {\mathcal{X}}}
         a_{ij}(k) \left(\overline{x}(k) - \varepsilon(k)\right)
        \nonumber \\
   & \le \overline x(k) - \gamma \varepsilon(k)
     = \overline x(k) - \varepsilon(k+1),
\end{align}
where the equality follows from \eqref{eqn:epsilon}.
This bound indicates that at the beginning of round $k+1$,
agent~$i$ will be outside $\overline {\mathcal{X}}(k,k+1)$.

On the other hand, we can show that each regular agent $i$
outside $\overline {\mathcal{X}}(k,k)$
will not go inside $\overline {\mathcal{X}}(k,k+1)$ at round $k+1$.
This is because it holds $x_i(k) \le \overline x(k) - \varepsilon(k)$
and $a_{ii}(k) \ge \gamma$. Hence, we can guarantee
the upper bound \eqref{eq-10} from \eqref{eq-09}.

Similarly, if regular agent~$i$ is
in $\underline {\mathcal{X}}(k,k)$ having $4f+1$
neighbors from outside the set
or is outside $\underline {\mathcal{X}}(k,k)$, then
we can lower bound its state as
\begin{equation}
  x_i(k+1)
    \geq \underline x(k) + \gamma \varepsilon(k)
    = \underline x(k) + \varepsilon(k+1).
\label{eq-10a}
\end{equation}
This indicates that such agent~$i$ will be outside of
$\underline {\mathcal{X}}(k,k+1)$ at the beginning of
round $k+1$.

Next, we discuss the updates of the cured agents with $\theta_i(k)=2$.
Take a cured
agent~$i\in\mathcal{C}(k)\cap \overline {\mathcal{X}}(k,k)$
having at least $4f+1$ neighbors outside $\overline {\mathcal{X}}(k,k)$.
Such an agent applies the deleting rule in Step~3\,(b2),
where its own value is removed as
\begin{align}
 x_i(k+1)
  &= \sum\limits_{j \in \mathcal{M}_i(k) \cap \overline {\mathcal{X}}}
      a_{ij}(k) x_j(k) + \sum\limits_{j \in \mathcal{M}_i(k) \setminus \overline {\mathcal{X}}}
      a_{ij}(k) x_j(k)
\label{eq-12}
\end{align}
with the self-weight $a_{ii}(k)=0$.
%The update rule depends on whether agent $i$ deletes its own value or not.
There are at most $2f$ values missing since each cured agent
with $\theta_i(k)=1,2$
does not send its value and removes at most $2f$ agents
outside $\overline{\mathcal{X}}(k,k)$.
So the set $\mathcal{M}_i(k)\setminus\overline{\mathcal{X}}(k,k)$
is guaranteed to be non-empty.
Then, we have from \eqref{eq-12} and then \eqref{eqn:epsilon}
\begin{flalign}
  x_i(k+1)
    \le \overline x(k) - \gamma \varepsilon(k)
      = \overline x(k) - \varepsilon(k+1).
\label{eqn:2A:cured}
\end{flalign}
Thus, cured agent~$i$ will be outside
$\overline{\mathcal{X}}(k,k+1)$ in the next round.
%we have that at least $f+1$ regular or cured agents
%at round $k$ move outside
%$\overline {\mathcal{X}} \left( {k,k{\rm{ + }}1,{\varepsilon}\left( {k + 1} \right)} \right)$.

It follows from \eqref{eq-10}, \eqref{eq-10a}, and
\eqref{eqn:2A:cured} that at round $k+1$,
the regular and cured agents in $\overline{\mathcal{X}}(k,k)$
or $\underline{\mathcal{X}}(k,k)$ having at least $4f+1$
neighbors from outside the corresponding sets
will be outside of both $\overline{\mathcal{X}}(k,k+1)$
and $\underline {\mathcal{X}}(k,k +1)$; the number of
such agents is $f+1$ or larger due case~3 considered so far.

\begin{figure}[t]
\centering\includegraphics[width=.98\linewidth]{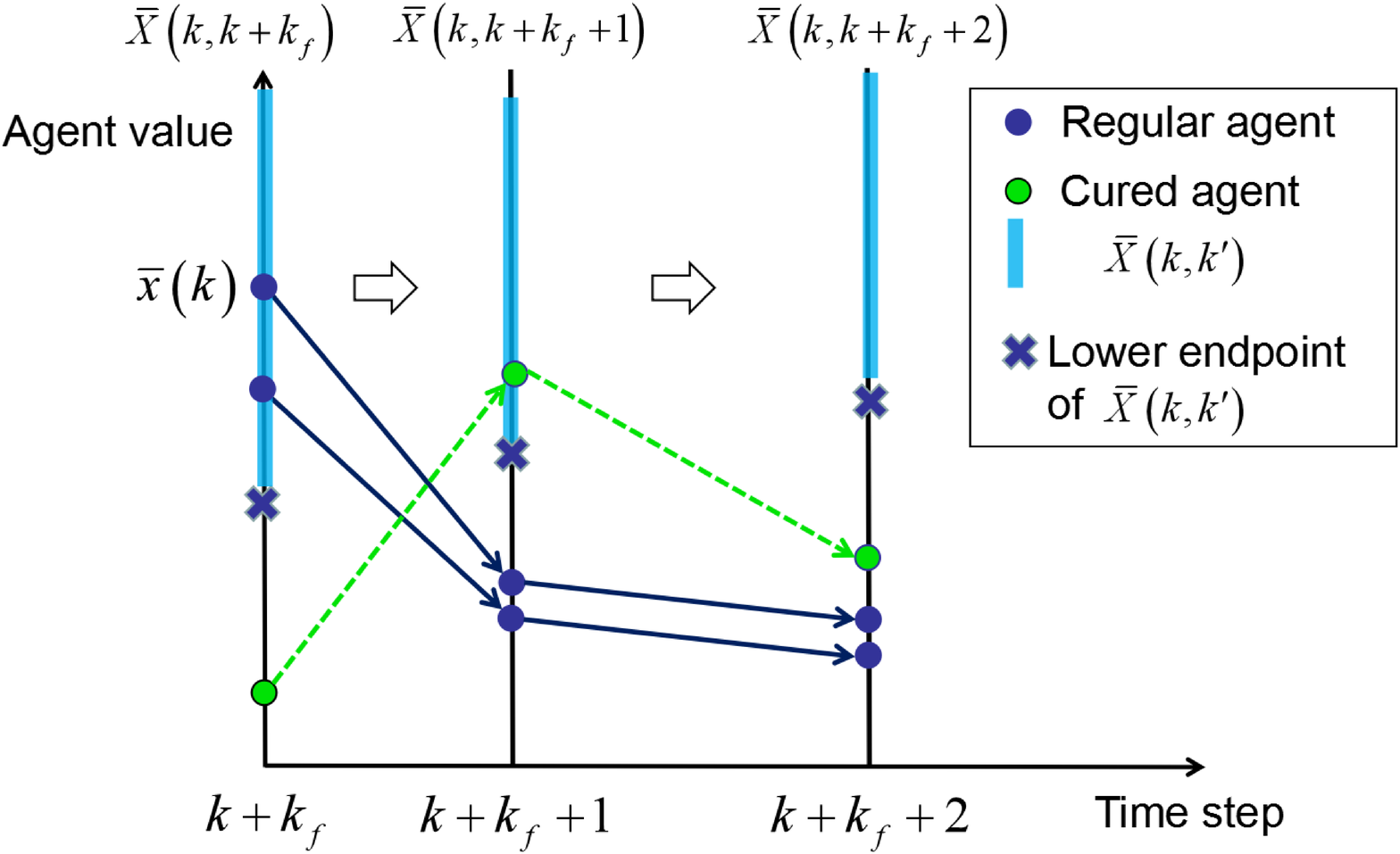}
\vspace*{-2mm}
\caption{Updates for regular and cured agents in case 1}
\label{fig06}
%\vspace*{-2mm}
\end{figure}

We now discuss the behavior of the cured agents (in $\mathcal{C}(k)$)
outside of both $\overline{\mathcal{X}}(k,k)$ and
$\underline{\mathcal{X}}(k,k)$.
Note that there are at most $f$~cured agents.
For agent $i \in \mathcal{C}(k) \setminus \overline{\mathcal{X}}(k,k)$,
the update rule is \eqref{eq-12},
%For update rule \eqref{eq-11}, we have similar argument and
%then we have \eqref{eq-10} for such class of cured agents.
%For update rule \eqref{eq-12}, we note that
and the updated value $x_i(k+1)$
may be inside $\overline {\mathcal{X}}(k,k+1)$;
this happens, for example, if the set
$\mathcal{M}_i(k) \setminus \overline{\mathcal{X}}(k,k)$
is empty.
Similar results hold for
agent~$i \in \mathcal{C}(k) \setminus \underline{\mathcal{X}}(k,k)$.
Thus, it follows that at most $f$~cured agents can
move inside $\overline {\mathcal{X}}(k,k+1)$ or
$\underline {\mathcal{X}} (k,k+1)$.

We summarize the arguments so far. During round $k$,
among the regular and cured agents
in $\overline {\mathcal{X}}(k,k+1)$ or $\underline {\mathcal{X}} (k,k+1)$
having $4f+1$ or more links from outside the corresponding sets,
at least $f+1$ of them move outside,
and at most $f$ cured agents might move inside these sets.
Hence, at least by one, the total number of such regular and cured agents in
%$\underline{\mathcal{X}}(k,k+1)$ is smaller than that
%in $\underline{\mathcal{X}}(k,k)$.
$\overline {\mathcal{X}}(k,k+1)$ and $\underline {\mathcal{X}} (k,k+1)$
is smaller than that in
$\overline {\mathcal{X}}(k,k)$ or $\underline {\mathcal{X}} (k,k)$.
Repeating this process, we eventually have that
there is some round $k+k_f$ such that the total number of
regular agents inside $\overline {\mathcal{X}}(k,k+k_f)$ and
$\underline {\mathcal{X}}(k,k+k_f)$ is smaller than $f+1$,
where $k_f>0$ is a finite number.
As a consequence, we know from the analysis above that
among the three cases 1--3 due to the
graph robustness mentioned above,
only case~1 and/or case~2 holds at round $k'=k+k_f$.

\if0
Note that at round $k+k_f$, we can check that
$\overline {\mathcal{X}} (k,k+k_f)$ and
$\underline {\mathcal{X}}(k,k+k_f)$ are
disjoint and nonempty.
By applying the graph robustness to these two sets yields
the following three cases:
\begin{enumerate}
  \item All agents in $\overline {\mathcal{X}}(k,k+k_f)$
        have at least $4f+1$ neighbors from outside.
  \item All agents in $\underline {\mathcal{X}}(k,k+k_f)$
        have at least $4F+1$ neighbors from outside.
  \item The total number of agents in $\overline {\mathcal{X}}(k,k+k_f)$
        and $\underline {\mathcal{X}}(k,k+k_f)$ that have at least $4f+1$
        neighbors outside is no smaller than $2f+1$.
\end{enumerate}
However, from the analysis above, we know that
only cases~1 and~2 are relevant and case~3 does not take place.
%we know that there are less than
%$f+1$ regular and cured agents inside $\overline {\mathcal{X}} \left( {k,k{\rm{ + }} k_f,{\varepsilon}\left( {k + k_f} \right)} \right)$ or $\underline {\mathcal{X}} \left( {k,k{\rm{ + }} k_f,{\varepsilon}\left( {k + k_f} \right)} \right)$ at time $k+k_f$. So case 3) cannot be satisfied any more. Case 1) and/or case 2) has to be satisfied.
\fi

Suppose that case~1 is satisfied (see Fig.~\ref{fig06}).
We show that at the end of round $k + k_f +2$,
the set $\overline {\mathcal{X}}(k,k + k_f+2)$
will not contain any regular agent or cured agent.
After the updates in round $k + k_f$,
we know from the analysis above that all regular and cured agents
inside $\overline {\mathcal{X}} (k,k + k_f)$ are outside the set
$\overline {\mathcal{X}} (k,k + k_f+1)$.
Moreover, at most $f$ cured agents in $\mathcal{C}(k+k_f)$ can be
inside $\overline {\mathcal{X}} (k,k + k_f+1)$
after the updates at $k+k_f$. Note that at the next round $k+k_f+1$,
such cured agents are in cured status with $\theta_i(k)=2$;
thus, they still do not send their values to neighbors, but make
updates as regular agents.
As a consequence, at round $k + k_f +1$, there is no value sent from agents
in the set $\overline {\mathcal{X}}(k,k + k_f+1)$.
This means that the regular agents outside this set will not move inside
$\overline {\mathcal{X}}(k,k + k_f+2)$ in the next update.
In the meantime, the cured agents in $\overline {\mathcal{X}}(k,k + k_f+1)$
move outside $\overline {\mathcal{X}}(k,k + k_f+2)$.
It thus follows that at round $k+ k_f +2$, all regular agents
and cured agents are outside $\overline {\mathcal{X}}(k,k + k_f+2)$.
Similar arguments also hold for agents in the set
$\underline {\mathcal{X}}(k,k+ k_f)$.
Therefore, at the end of round $k + k_f+ 2$, at least,
one of the two sets $\overline {\mathcal{X}}(k,k + k_f+2)$ and
$\underline {\mathcal{X}}(k,k+ k_f+2)$
is empty of regular and cured agents.

First, consider the case for $\overline {\mathcal{X}}(k,k + k_f+2)$
containing no regular/cured agents.
Then, we have for all $i \in \mathcal{R}(k)$.
\[
  x_i(k + k_f +2)
   \le \overline{x}(k) - \gamma ^{k_f+2}\varepsilon(k).
\]
It thus follows that
\[
  \overline{x}(k + k_f + 2)
   \le \overline{x}(k)
         - \gamma^{k_f+2} \varepsilon(k).
\]
Recall that $\overline{x}(k)$ is nonincreasing and
$\underline{x}(k)$ is nondecreasing based on the
update rule \eqref{eqn:msr}.
Hence,
\begin{flalign}
  &V(k +k_f +2)
    = \overline{x}(k +k_f +2)
         - \underline{x}(k +k_f +2)\nonumber\\
   &~~~~~\le \overline{x}(k)
         - \gamma ^{k_f+2} \varepsilon(k)
         - \underline{x}(k) %\nonumber\\
    = \left( 1- \frac{\gamma^{k_f+2}}{2} \right) V(k).
 \label{eqn:Vkf}
\end{flalign}
Note that the analysis is similar
for the other case where
$\underline {\mathcal{X}}(k,k+ k_f+2)$ is empty of regular/cured agents.
That is, the bound in \eqref{eqn:Vkf} holds in either case.
Repeating this argument, we have
\[
  V(k + l (k_f +2))
   \le \left( 1- \frac{\gamma ^{k_f+2}}{2} \right)^{l} V(k).
\]
Therefore, we have $V(k) \to 0$ as $k \to \infty$ and thus
the consensus condition holds.
$\hfill\square$

\smallskip
In Theorem~\ref{theorem09}, the sufficient condition for
resilient consensus is expressed in terms of
the graph condition based on the notion of robustness.
The analysis follows approaches employed in the recent literature
on MSR algorithms for static malicious models
(e.g., \cite{LeBlanc2013, Dibaji2015}).
Compared with conventional MSR algorithms,
the proof techniques are different in mainly three aspects:

(i) Not only the adversary agents in the graph send corrupted
values, but also the cured agents exhibit non-regular behaviors
by not sending values to neighbors.
Furthermore, the cured agents in $\mathcal{C}(k)$ follow
an update rule different from the regular agents.
In the proof, we have to separately analyze the updates
of such cured agents.

(ii) The behavior of the cured agents in $\mathcal{C}(k)$ is
unique as they do not obey the normal update rules for regular agents.
It is interesting that the cured agents outside
$\overline {\mathcal{X}}(k,k)$ may move inside the
set $\overline {\mathcal{X}} (k,k+1)$.
Such responses do not occur in the analysis of conventional MSR
algorithms.

(iii) In the proof of Theorem~\ref{theorem09},
the set $\overline {\mathcal{X}}(k,k+k_f+2)$
is shown to be empty after round $k+k_f+2$.
The extra two rounds after $k+k_f$ are needed, again, because
of the behavior of cured agents.
We have shown that in the worst case,
some cured agents may first move inside the set
$\overline {\mathcal{X}}(k,k+k_f+1)$ and then move outside
$\overline {\mathcal{X}}(k,k+k_f+2)$.
In our result, we have to guarantee that this set is empty
of both regular and cured agents, while in conventional
studies, this is needed only for regular agents.

\begin{figure}[t]
\centering
\includegraphics[width=1\linewidth]{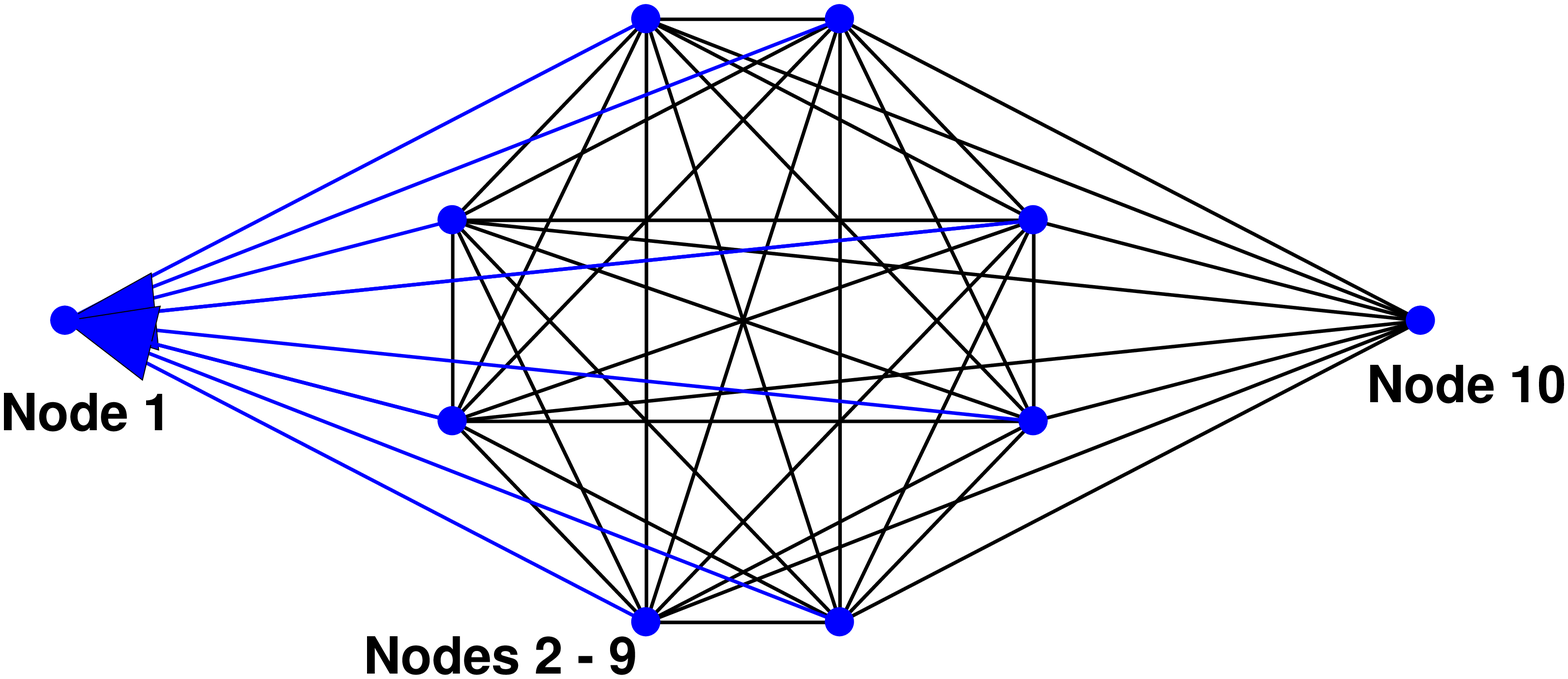}
\vspace*{-4mm}
\caption{Example of a $(5,3)$-robust graph}
\label{fig.add}
%\vspace*{-2mm}
\end{figure}

\renewcommand{\arraystretch}{1.05}
\begin{table*}[t]
\centering
\caption{Properties of adversary models and algorithms}
\vspace*{-2mm}
\begin{tabular}{c|cccccc}
\hline
Adversary &   Algorithm &   \# deleted & \# cured values
%            & Necessary \# & Network condition \\
            & Network condition \\
model     &             &   values & not transmitted
%            & of neighbors & (with $s\geq 0$)    \\ \hline
            & (with $s\geq 0$)    \\ \hline
Static model       &   Conventional MSR
 &      From $0$ to $2f$          &           0 & $(f+1,f+1)$-robust \\
M1                 &   Protocol 1
 &      $2f$                      &           0 & Part of $(2f+1,s)$-robust\\
M2                 &   Protocol 2
 &      $2f$                      &   From $0$ to $f$ & Part of $(3f+1,s)$-robust  \\
M2                 &   Protocol~2A
 &      From $0$ to $2f$          &   From $0$ to $2f$ &$(4f+1,2f+1)$-robust  \\
M3                 &   Protocol 3
 &      $4f$                      &           0 &  Part of $(4f+1,s)$-robust \\ \hline
\end{tabular}
\label{tab2}
%\vspace*{-2mm}
\end{table*}

\subsection{Discussion}

We now discuss the relations among the graph conditions that
appeared in our results in this paper.

\subsubsection{Relation between Protocols~2 and 2A}
%First, the relation between Protocols~2 and 2A.
Compared with Protocol~2,
the main difference of Protocol~2A is the
relaxed deleting rules applied to the regular agents,
allowing them to use the safe values are more efficiently.
More specifically, as the cured agents refrain from sending
their values for two consecutive rounds,
among the values of the neighbors received by each regular agent,
the ratio of safe and reliable ones sent from regular agents
is higher.
As a result, Protocol~2A can guarantee resilient consensus
for more sparse networks in comparison with
the non-complete graph case for Protocol~2.
In particular, for a fixed number $f$ of malicious agents,
as the network size becomes larger,
the connectivity condition for Protocol~2A in Theorem~\ref{theorem09}
may become less than that for Protocol~2 in Theorem~\ref{theorem04};
recall that the latter result requires
every agent to have at least $n/2$ neighbors.
Moreover, the graph condition for Protocol~2A is determined
only by $f$.

We demonstrate the differences between Protocols~2 and~2A through two
examples. The first is related to the graph conditions in the two
theorems. Consider networks with ten agents ($n=10$)
with one mobile malicious agent ($f=1$).
It is easy to check that in this case,
the conditions for Protocol~2 in Theorem~\ref{theorem04} are satisfied
only under the complete graph since the required number of
neighbors for each agent is $3f+1+n/2=9$.
On the other hand,
the non-complete graph shown in Fig.~\ref{fig.add} with ten
nodes is $(5,3)$-robust and thus satisfies the condition with $f=1$
for Protocol~2A in Theorem~\ref{theorem09}.
In this graph, nodes 2--10 form a clique (i.e., a complete subgraph),
but among them, only nodes 2--9 have
directed edges (in blue) towards
agent~1. Note that all agents have only eight (incoming) neighbors.

\if0
\renewcommand{\arraystretch}{1.05}
\begin{table}[t]
\centering
\caption{Properties of the mobile adversary models}
\vspace*{-2mm}
\begin{tabular}{c|ccccc}
\hline
%       & \multicolumn{2}{|c|}{Event-Triggered}
%       & Time- \\ \cline{2-3}
Model              &   Algorithm             & Network condition                   \\ \hline
Static model       &   Conventional MSR      &      $(f+1,f+1)$-robust          \\
M1                 &   Protocol 1            &      $(2f+1,s)$-robust  \\
M2                 &   Protocol 2            &      $(3f+1,s)$-robust  \\
M2                 &   Protocol~2A            &     $(4f+1,2f+1)$-robust       \\
M3                 &   Protocol 3            &      $(4f+1,s)$-robust  \\ \hline
\end{tabular}
\label{tab2}
%\vspace*{-2mm}
\end{table}
\fi

%\begin{figure*}[t]
\begin{figure}[t]
  \centering
      \includegraphics[width=1\linewidth]{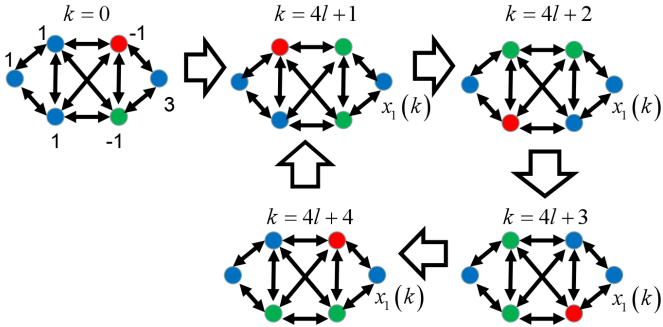}
    \vspace*{-.5cm}
    \caption{An illustration of Protocol~2A with one mobile malicious agent
             with time index $l\geq 0$}
      \label{fig.p4}
%\end{figure*}
\end{figure}

In the next example, we illustrate how Protocol~2A
can achieve resilient consensus while Protocol~2 cannot
because of the difference in their update rules.
Here, we consider the graph shown in Fig~\ref{fig.p4}
with six agents with one mobile malicious node (i.e., $f=1$).
For our purpose, a simple graph is taken,
which does not satisfy the theoretical conditions.
The graph has two nodes having two neighbors
and four nodes having four neighbors;
among the latter four nodes, the malicious agent
moves around periodically, following the M2 model.
Note that the regular agents take only positive values while
the malicious agent always takes $-1$, which results
in the cured agent with $\theta_i(k)=1$ take the same negative value.

At the initial step, one malicious agent and one
cured agent are present indicated by red and green, respectively.
In Fig~\ref{fig.p4}, the left-most plot shows the status
of the agents and their values chosen as $[1~1~1~-1~-1~3]$
at the start of round $k=0$. Note the only agent not in
consensus at this time is the agent on the far right of
the graph taking the initial value $3$; this agent will
be called agent~1 with value $x_1(0)=3$.
After the initial round,
as shown in Fig~\ref{fig.p4} on the right side, the malicious agent
moves periodically among four agents in this network.
Under Protocol~2, the updates will fail even at the initial
round. With $f=1$, agent~1
cannot update its value since it must have at least
four values from its neighbors.

On the other hand, in Protocol~2A, each regular agent employs
an update rule based on the conventional MSR algorithm.
It hence always keeps and uses its own value in the updates.
Thus, at round $k=0$, agent~1 keeps its value unchanged as $x_1(1)=3$,
removing the value $-1$ received from the malicious agent.
In fact, it remains unchanged in the following two rounds
as $x_1(2)=x_1(3)=3$,
because it does not receive enough values because of
the cured neighbors. At round $k=3$, for the
first time, agent~1's value changes to $x_1(4)=2$
by taking average of $1$, $-1$, and $3$.
by taking average of $1$ and $3$
since the value $-1$ is discarded by the protocol.
Due to the periodic change in the
values, we can show that agent~1 will update
its value at rounds $k=4\ell$
as $x_1(4(\ell+1))=(x_1(4\ell)+1)/2$ for $\ell=0,1,\ldots$.
while the values of other regular agents remain at 1.
Clearly, it holds $x_1(k)\to 1$ as $k\to\infty$,
and thus resilient consensus will be achieved.

%From this picture, we know that there are typically four steps
%of one round after the initial state. Observe the second step
%of each round, we can see that the value of the right agent is
%converging to the consensus value 1 as time goes by.
%When time goes to infinity, we can see that the resilient consensus
%is achieved.

%Next, we see how the update rules as well as the behavior
%of the cured nodes can affect the performance of the two protocols.

\subsubsection{Network Conditions and Their Robustness}
Next, we would like to relate the graph conditions
obtained in Theorems~\ref{theorem02}, \ref{theorem04},
and \ref{theorem06} with robust graphs.
The following result provides the means to do so.

\begin{proposition} \label{theorem10}
For a given graph $\mathcal{G}$ and a nonnegative integer $r$,
if the number of neighbors for each node~$i$
satisfies $|\mathcal{N}_i| \ge r + n/2$, then this
graph is $(r,s)$-robust, where $s$ can be an arbitrary
nonnegative integer.
\end{proposition}

%\smallskip
\noindent
\textit{Proof}:
At first, we take two disjoint subsets $\mathcal{S}_1$
and $\mathcal{S}_2$ of the node set $\mathcal{V}$
with $|\mathcal{S}_1| \le |\mathcal{S}_2|$.
There are two cases:
%1). $\mathcal{S}_1$ contains more than $n/2$ agents and
%$\mathcal{S}_2$ contains less than $n/2$ agents.
(i) $\mathcal{S}_1$ contains less than or equal to $n/2$ agents, and
$\mathcal{S}_2$ contains no fewer than $n/2$ agents.
(ii) Both $\mathcal{S}_1$ and $\mathcal{S}_2$ contain
less than $n/2$ agents.
It is noted that one of these sets must contain less than or
equal to $n/2$ agents.
Then, we check every agent~$i$ in $\mathcal{S}_1$.
Since the number of its neighbors satisfies $|\mathcal{N}_i| \ge r + n/2$,
we easily see that every agent inside $\mathcal{S}_1$
must have at least $r$ neighbors from outside $\mathcal{S}_1$.
Hence, the condition 1 or 2 in Definition~\ref{robust_graph}
must be satisfied. Since we do not need to check condition~3,
the parameter $s$ can be chosen arbitrarily.
$\hfill\square$

\smallskip
This result demonstrates that the graph conditions of
Theorems~\ref{theorem02}, \ref{theorem04}, and~\ref{theorem06}
can be stated in terms of robust graphs.
We know that checking the robustness of large graphs is
combinatorial and thus challenging. This proposition
provides a simple analytic method to design networks
with robustness properties.

Table~\ref{tab2} summarizes the properties of the five
protocols under the four adversary models discussed in this paper
including the conventional MSR algorithm under the static model.
As shown in the table, for example, graphs satisfying the conditions
for Protocol~1 in Theorem~\ref{theorem02} have the property of
$(2f+1,s)$-robustness with any $s\geq 0$. Note however
that this is only a necessary condition,
and the converse does not hold in general.
That is, in $(r,s)$-robust graphs, each node~$i$ does not
necessarily satisfy $|\mathcal{N}_i| \ge r + n/2$.
Further, from Table~\ref{tab2}, we can find that as the mobile adversarial
model becomes more powerful, the required connectivity level
also increases. This table also gives a comparison between
the conventional MSR for static models and our proposed protocols
for mobile models.

\section{Numerical Example}
\label{Section 7}

In this section, we illustrate the performance of our proposed
protocols and the conventional MSR algorithm
under mobile adversary models
through a numerical example using a wireless multi-agent network.

Our focus of the numerical experiments is to determine how well
the protocols perform under practical settings when the assumptions
introduced in the theoretical development may not hold.
Specifically, we use randomly generated networks where the connectivity
requirements are in general difficult to check due to the size of the
network.
Furthermore, we consider uncertain situations regarding the information
of the adversarial agents in terms of their models and numbers.
To this end, we use random graphs with 100~nodes and change the connectivity
levels to examine the success rates for achieving resilient consensus
through extensive simulations. We also check the cases where
the parameter $f$ for the number of adversaries may be smaller
than the actual number of such nodes; the latter number is
denoted by $f_{\text{real}}$ in this section.

\begin{figure}[t]
\vspace*{-2mm}
\centering
\includegraphics[width=8cm,height=7cm]{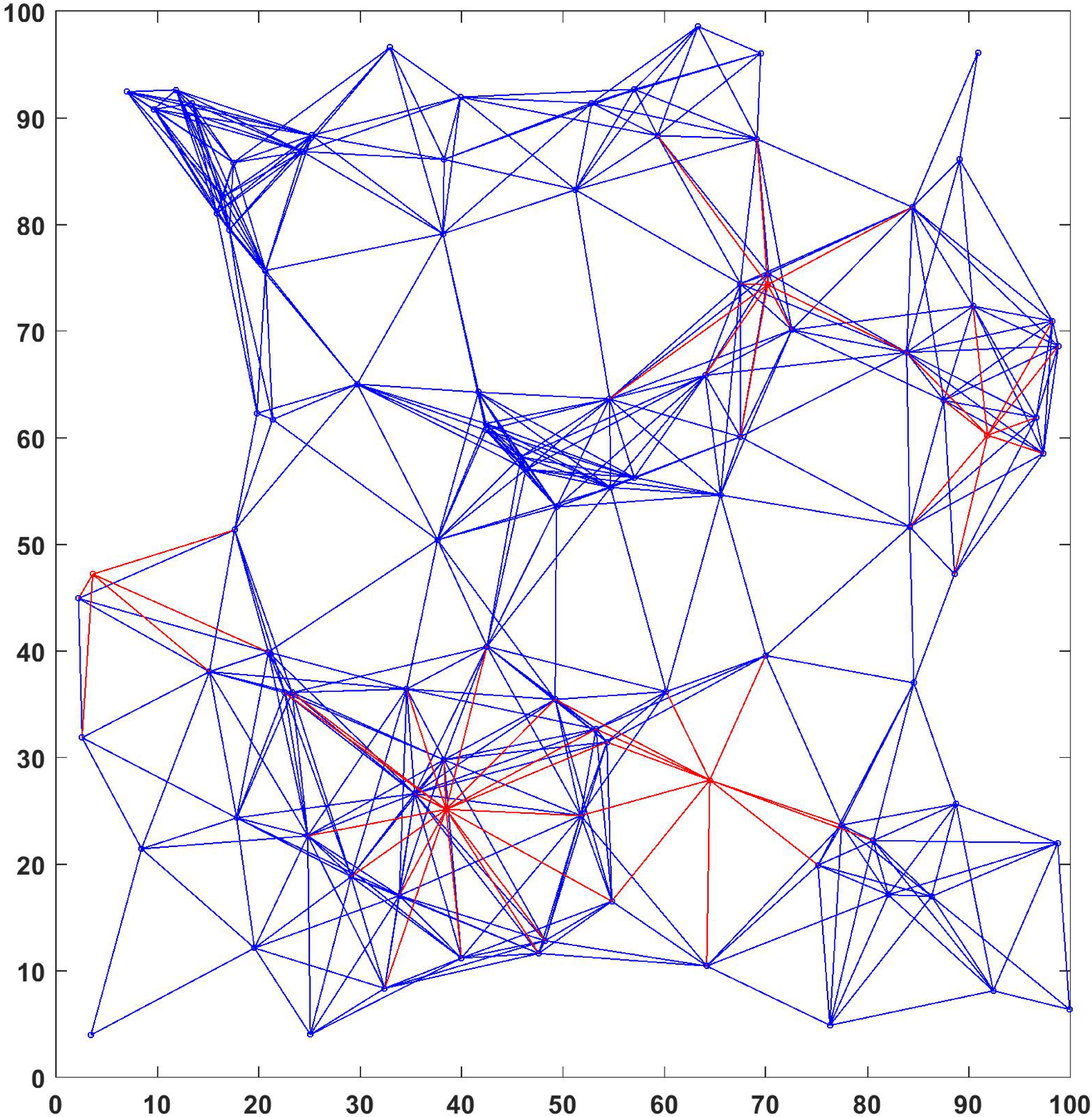}
\vspace*{-4mm}
\caption{Wireless multi-agent network with 100 nodes ($r=20$)}
\label{fig.g1}
\end{figure}

For the network topologies, we generated ten random geometric graphs with
100 nodes located in an area of $100$ meters
$\times$ $100$ meters randomly under the uniform distribution.
Each agent has a communication range
determined by the radius $r$, within which
it can communicate with all agents.
An example is shown in Fig.~\ref{fig.g1}
where the communication radius is chosen as $r=20$.
The regular agents and their edges are drawn in blue while
the malicious agents are in red.
Here, we placed 5 malicious agents, that is, $f_{\text{real}}=5$.

\begin{figure}[t]
\centering
%\vspace*{-1mm}
\includegraphics[width=1\linewidth]{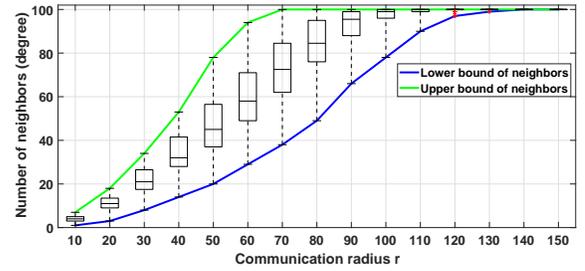}
\vspace*{-4mm}
\caption{Average number of neighbors versus the communication radius $r$}
\label{fig.g2}
\end{figure}

In the box plot of Fig.~\ref{fig.g2},
we display the distribution of the number
of each agent's neighbors for the topology in Fig.~\ref{fig.g1}
versus the radius $r$. For each $r$,
the green and blue curves indicate the maximum and the minimum
numbers of neighbors, respectively, while
the box represents the range containing the first to third quartiles
and the line in the box shows the median.
For large values of $r$, a few red crosses are shown,
indicating outliers.

In this experiment, we ran the algorithms under three settings
to examine the success rates for resilient consensus.
Throughout the simulations,
the regular nodes' initial values were randomly chosen under uniform
distribution in the interval $[0,100]$.
On the other hand, the adversary nodes were given negative values so that
their influence is easy to see.
For the mobile adversaries,
we used the random model, under which at each time step,
the malicious agents randomly choose nodes to move from the
entire network.
%\item (\emph{Maximum value mobile model}) The adversary node randomly chooses a regular node taking the maximum value at each time step.

\begin{figure*}[t]
  \centering
  \vspace*{-2mm}
  \includegraphics[width=.97 \linewidth]{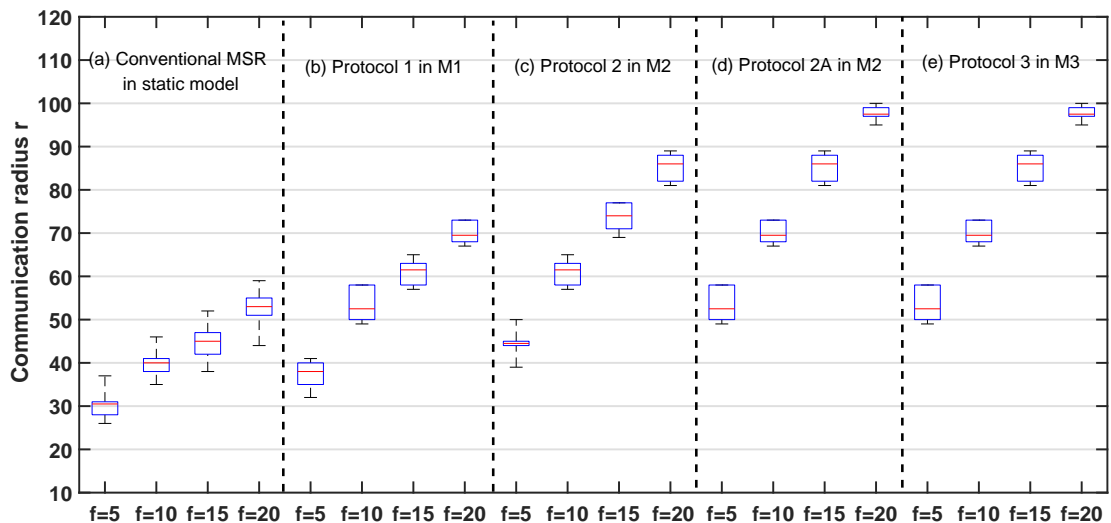}
  \vspace*{-4mm}
  \caption{Distribution of the threshold radii under five attack scenarios}
  \label{fig.g3_}
%  \vspace*{-2mm}
\end{figure*}

\subsection{Consensus under Different Communication Radii}
\label{sec:numerical:1}
As we have observed in the theoretical results, the different models
in the adversaries require different levels of connectivities in
the network. In the first part of the simulations, we verify such
properties of the algorithms by
finding the smallest communication radius over which resilient
consensus becomes possible. Such a communication radius will be
referred to as the \textit{threshold} radius.

\begin{figure*}[htb]
  \centering
%  \vspace*{-1mm}
  \subfigure[Conventional MSR in static model]{%
      \label{fig.g8}
      \includegraphics[width=.48\linewidth]{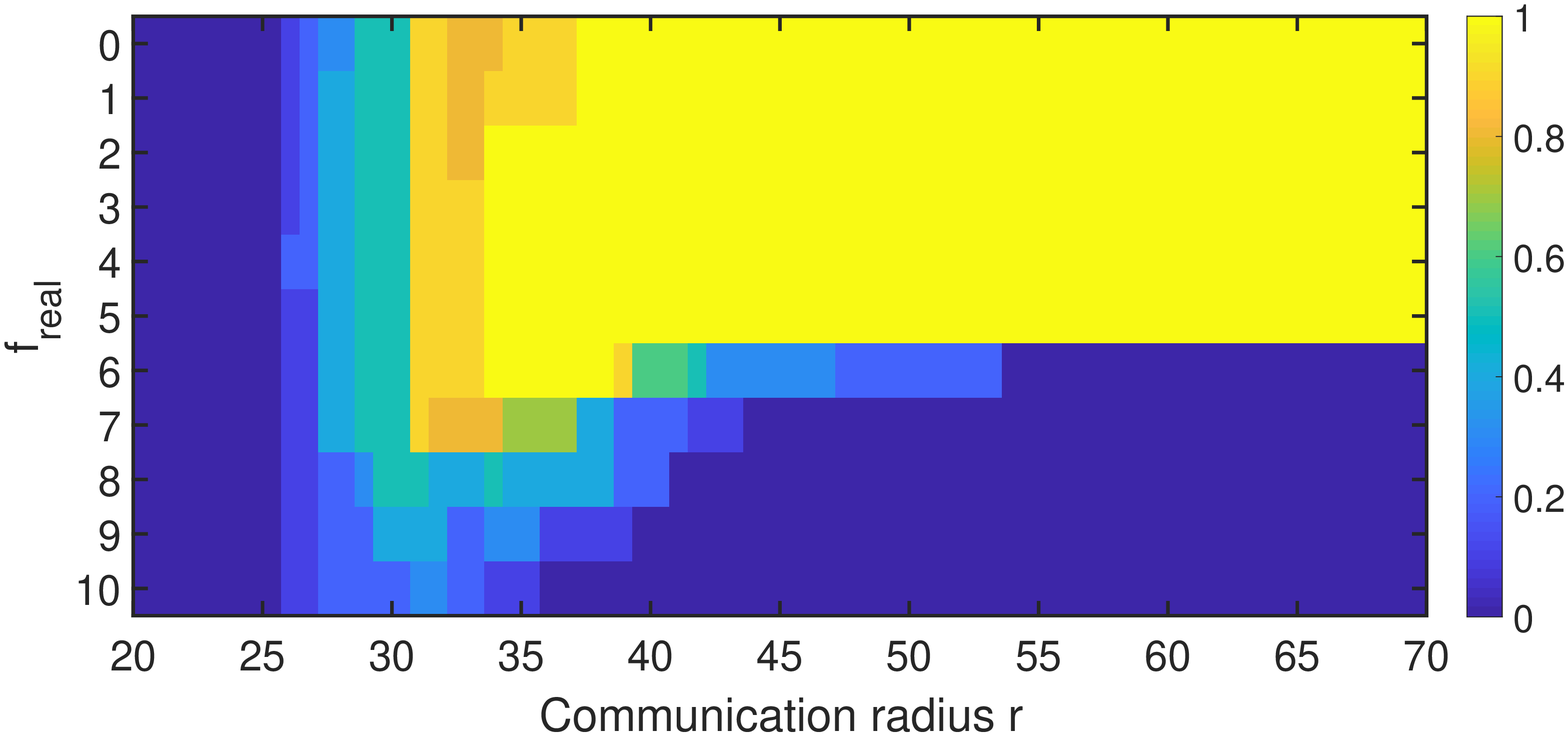}
      \vspace*{-2cm}}
  \subfigure[Protocol 1 in M1]{%
       \label{fig.g9}
           \includegraphics[width=.48\linewidth]{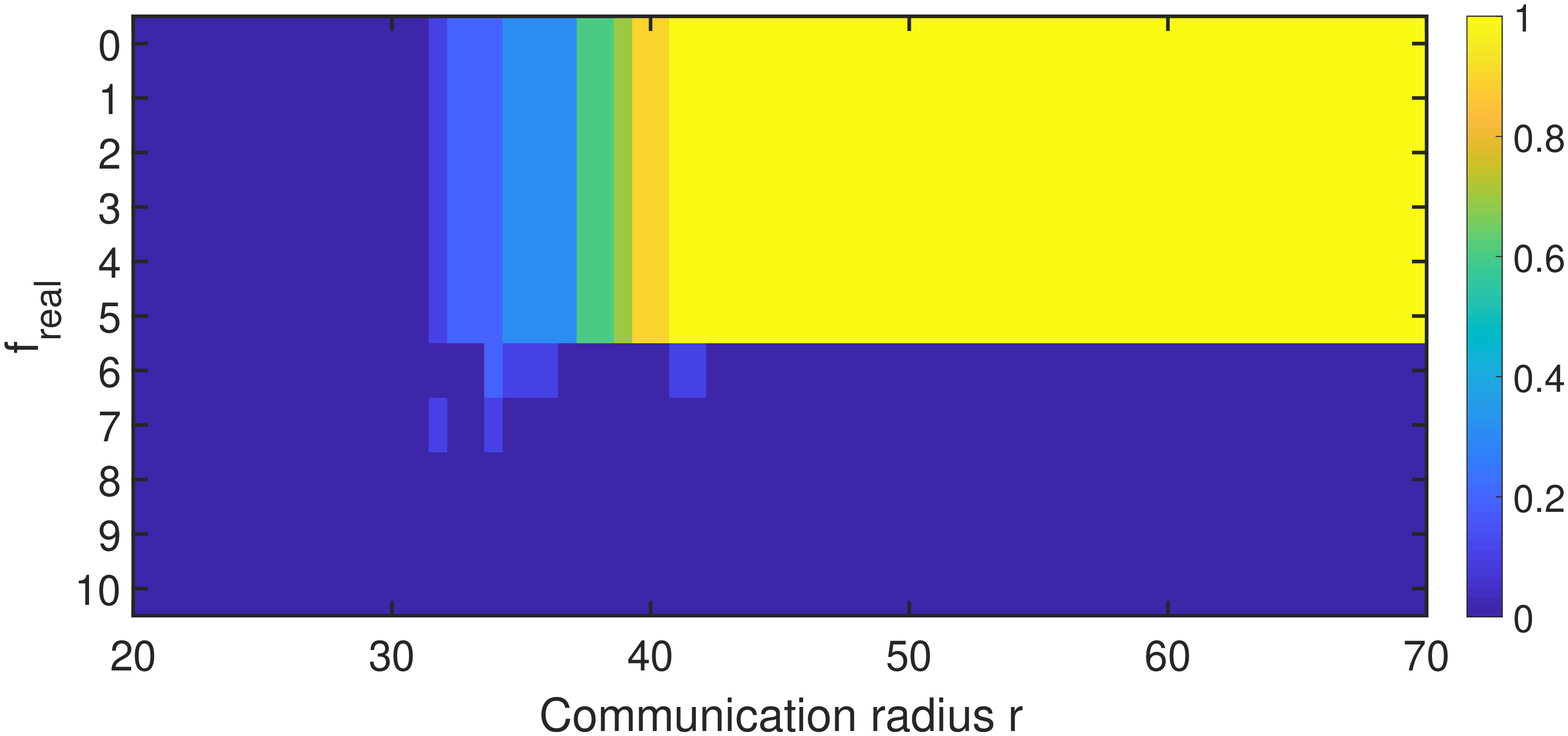}
           \vspace*{-2cm}}
  \subfigure[Protocol 2 in M2]{%
      \label{fig.g10}
      \includegraphics[width=.48\linewidth]{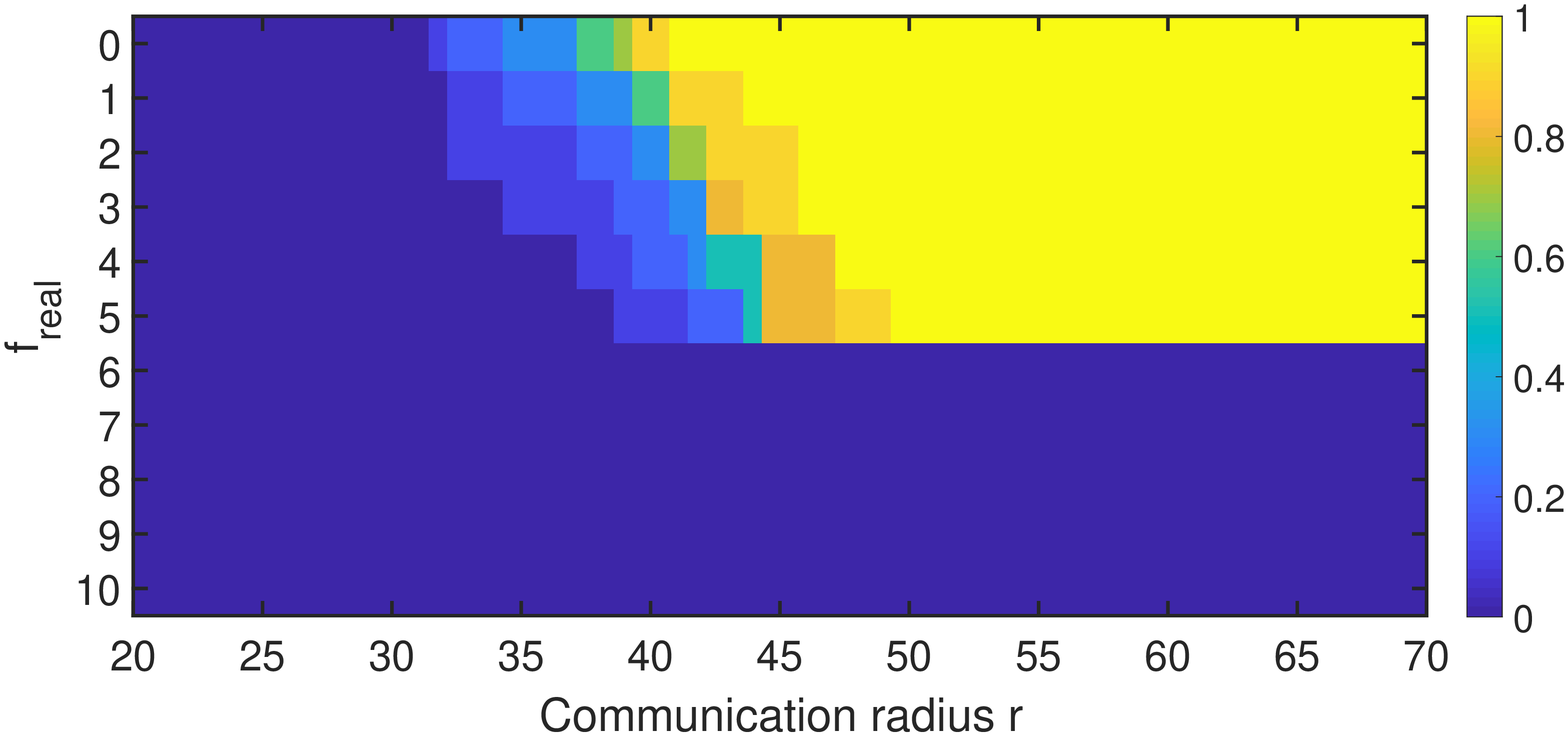}
      \vspace*{-2cm}}
  \subfigure[Protocol~2A in M2]{%
           \label{fig.g12}
           \includegraphics[width=.48\linewidth]{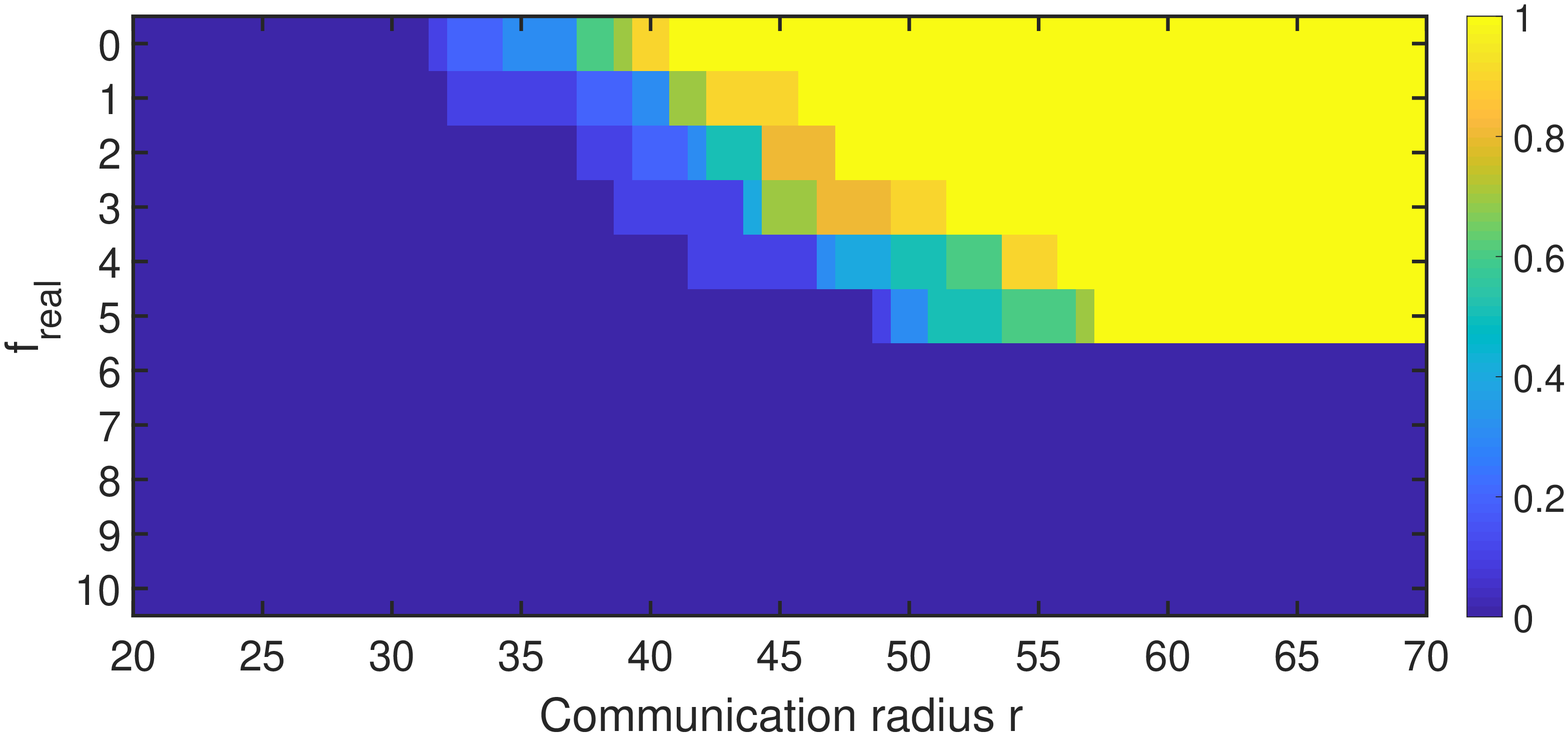}
           \vspace*{-2cm}}
  \subfigure[Protocol 3 in M3]{%
           \label{fig.g11}
           \includegraphics[width=.48\linewidth]{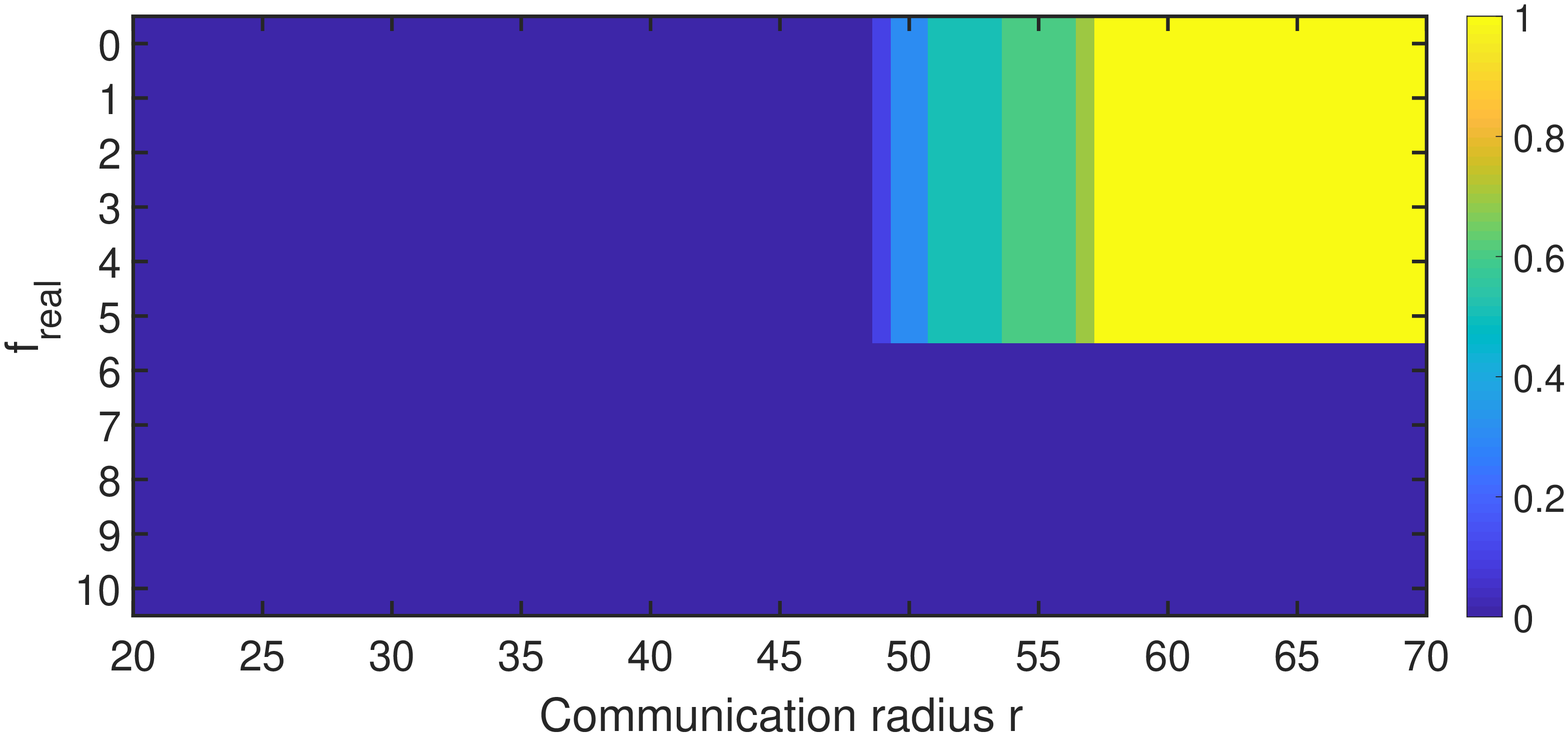}
           \vspace*{-2cm}}
%  \vspace*{-1mm}
  \caption{Success rates of resilient consensus
           versus the actual number $f_{\text{real}}$ of malicious agents
           and $r$ in the regular agents with parameter $f=5$}
 \label{fig.ggg}
%  \vspace*{-2mm}
\end{figure*}

In the simulations, we examined the five cases as follows:
(a)~Conventional MSR protocol under the static model,
(b)~Protocol~1 under the M1 model,
(c)~Protocol~2 under the M2 model,
(d)~Protocol~2A under the M2 model, and
(e)~Protocol~3 under the M3 model.
For each protocol, we checked four cases with the parameters $f=5,10,15,20$,
where the actual numbers of adversaries were taken as $f_{\text{real}}=f$.
The results are summarized in Fig.~\ref{fig.g3_}, where
the threshold radii are shown for the five
cases (a)--(e) from the left to right;
the figure is given in the box plot style
for the 10 network topologies generated as explained above.

We can observe two general trends in the results: First, for each
protocol, as the number $f$ of adversaries increases, the threshold
radius becomes larger. In the plots, the increase in the threshold size
seems to be linear in $f$ for all five protocols.
Second, the malicious nature in the models of the adversaries gradually
becomes higher in the following order: the static model, M1, M2, and M3.
Correspondingly, in the plots, we can confirm that from case (a) to case (d),
the required levels of the threshold radii become larger.

It is interesting that when we compare the cases (c) for Protocol~2
and (e) for Protocol~2A, which are both for the same model M2, the
thresholds are different. In fact, those for Protocol~2A are larger,
being comparable to those for Protocol~3 under M3.
%Note that in these simulations, Protocol~2A for M2 requires
%higher level of connectivity than Protocol 2 for M2 to guarantee
%resilient consensus.
This is because of the two cured rounds required in Protocol~2A since
these rounds lead the network to be less connected.
Recall that in Protocol~2A, the number of agents not sending their values
can be up to $2f$, while in Protocol~2, it is up to $f$.
More specifically, in Protocol~2A, the regular agents and the cured agents
with $\theta_i(k)=2$ remove less neighbor information than the regular
agents in Protocol~2. On the other hand, the cured agents in $\mathcal{C}(k)$
in Protocol~2A remove the same numbers of neighbors as those in Protocol~2.

We further note that the threshold results for cases (d) and (e)
exactly coincide though the adversary models as well as the protocols
are different. This is rather due to the special set up in the current
simulations. In case (d) for Protocol~3, each regular agent always
removes $4f$ values from those received from neighbors.
On the other hand, in case (e) under the model M3,
cured agents do not send their values.
Hence, since the malicious agents take negative values,
the regular agents employing Protocol~2A or~3
under the corresponding model are always successful in
ignoring the malicious agents' values, resulting in the same behaviors
and thus the same threshold values.

%The related success rate plots are shown in Fig . From these five plots, we can see that in static model, the conventional MSR algorithm needs the least connection. With communication radius 40, we can guarantee the resilient consensus almost surely when $f=5$.

%We confirm that from Protocol 1 for M1 to Protocol 3 for M3, as the mobile malicious model becomes more powerful, we need more connections in the graph.
%{Note that in these simulations, Protocol~2A for M2 requires higher level of connectivity than Protocol 2 for M2 to guarantee resilient consensus. This is because of two cured rounds, which lead to more disconnections in the network. Comparing Protocol 2 and Protocol~2A, we can see that the regular agents and the agents in $\mathcal{C}_2$ in Protocol~2A remove less neighbor information than the regular agents in Protocol 2. The agents in $\mathcal{C}_1$ in Protocol~2A removes the same number of neighbors as Protocol 2. However, the agents that do not send their values in Protocol~2A can be up to $2f$, while in Protocol 2, they are up to $f$.

%Note that in all mobile protocols, the random mobile model and maximum value mobile model have similar performance. Thus, in the following simulations, if there is no additional explanation, we consider the random mobile model since the related curves are also similar in maximum value mobile model.

\subsection{When the Number of Malicious Agents is Unknown}
\label{sec:numerical:2}

In the first part of the simulations, we have assumed that the number of
the malicious agents in the network is known, that is, $f=f_{\text{real}}$.
In this second part, we check the performance of proposed algorithms
when this does not hold, that is,
the parameter $f$ used in the algorithm does not match
the actual number $f_{\text{real}}$ of malicious agents.
Here, we fixed the parameter $f$ in the protocols at $f=5$ and
computed the success rates of resilient consensus by changing two parameters:
the number of real malicious agents as $f_{\text{real}}\in\{0,1,\ldots,10\}$
and the communication radius as $r\in[20,70]$.
In Fig.~\ref{fig.ggg},
the results are shown in the heat map format for the five cases (a)--(e)
as in the previous part.

These plots reveal the sharp difference between
the static and mobile malicious models.
Fig.~\ref{fig.g8} shows the results of the conventional MSR algorithm
under the static malicious model.
We can see that the success rate is quite low with $r \leq 30$ as
the connectivity in the network is not enough.
With $r > 30$, however, enough connectivity is introduced in the networks,
and resilient consensus can be guaranteed in most topologies
when $f_{\text{real}} \le f$.

Moreover, under the static malicious model, an interesting phenomenon
occurs when more malicious agents exist in the network with
$f_{\text{real}} > f$. Observe that
the success rate is high when $30 \leq r \leq 40$.
In this range, the communication radius is relatively small.
Low connectivity in the network actually helps the regular agents
since there are certain chances for each regular agent
to receive fewer than $f$ malicious values.
The conventional MSR is then able to achieve resilient consensus.

In contrast, under any of the mobile malicious models, this does not
happen, and it is critical to take the parameter $f$ large enough that
$f_{\text{real}} \leq f$ holds. In fact, in
Figs.~\ref{fig.g9}--\ref{fig.g12}, the success rates almost immediately go to 0
once $f_{\text{real}} > f$.
In the mobile models, the adversaries can switch among agents
in the network, so once $f_{\text{real}} > f$ holds, there is a large
chance that the regular agents receive more than $f$
malicious values at some time instants. After such moments,
resilient consensus becomes impossible to reach.
Thus, in mobile models, resilient consensus is very hard to guarantee
when $f_{\text{real}} > f$.

Furthermore, according to the malicious models, the sizes of the region
indicating high success rates are different.
As we have already seen in the previous part,
the success rate becomes lower as the model becomes more adversarial
from case (b) to case (e). It is interesting that under the model M2,
for both Protocols~2 and~2A in
Figs.~\ref{fig.g10} and Fig.~\ref{fig.g12},
respectively, the success rates are affected by the
size of $f$ even when the relation $f\geq f_{\text{real}}$ holds.
This is because depending on the size of $f_{\text{real}}$,
the number of cured agents is determined. Hence,
as $f_{\text{real}}$ increases, the overall system
is under more uncertainties and its performance becomes worse.
Note that even between these two protocols,
the decrease in the success rates is different.
Compared with the heat map for Protocol~2 in Fig.~\ref{fig.g10},
the one for Protocol~2A in Fig.~\ref{fig.g12}
indicates that the yellow region shrinks faster for larger
$f_{\text{real}}$. On the other hand,
in the results for Protocols~1 and~3 in Figs.~\ref{fig.g9} and
Fig.~\ref{fig.g11}, respectively,
the size of the parameter $f$ has
much less impact on the chances of achieving
resilient consensus. The reason is that in both protocols,
the regular agents always ignore $2f$ values at each update
and this number stays the same regardless of the values received.

\subsection{When the Mobile Malicious Model is Unknown}
\label{sec:numerical:3}
In the third part of the simulations, we introduce further uncertainties
in the setting. In addition to using
the actual number $f_{\text{real}}$ of malicious agents greater
than the parameter $f$ used in the algorithms,
we run the five protocols under different mobile models.
As in the previous part, we fix the parameter as $f=5$.
Then, for two sets of network topologies, namely, with
communication radius $r=70$ and $r=50$,
we performed simulations to calculate the success rates of resilient consensus
by changing $f_{\text{real}}$ from 0 to 10.
For each case, the maximum $f_{\text{real}}$ under which
resilient consensus is reached in all 10 topologies was recorded.
The results are displayed in Tables~\ref{tab:add1}\,(a) and
\ref{tab:add1}\,(b) for $r=70$ and $r=50$, respectively.

In what follows, we discuss the case of $r=70$ with more connectivities
among the agents for the four adversary models:

(i)~In the static model, all five protocols can achieve resilient consensus
under the 10 network topologies when the number of malicious agents in
the network is less than the bound, i.e., $f_{\text{real}}\leq f$.
Protocol~3 is special in that it can tolerate up to $2f=10$ malicious
agents;  this is because it ignores $2f$ largest and smallest neighbor values.

(ii)~For the mobile model M1, we can check that the conventional MSR fails to reach consensus
as soon as one malicious agent is introduced (with $f_{\text{real}}\geq 1$).
In the meantime, the proposed protocols perform well for this mobile
model when $f_{\text{real}}\leq f$.
Again, Protocol~3 can further tolerate up to $2f$ malicious agents.
Once the number $f_{\text{real}}$ exceeds the bound $f=5$
(or $2f=10$ for Protocol~3), all protocols fail to reach resilient consensus
for any of the topologies.

(iii)~For the mobile M2,
it is clear that the conventional MSR and Protocol~1 fail
to reach consensus even with one malicious agent in the network while
other protocols work well.
% when the communication radius is large enough with $r=70$.
Note that in this mobile model, Protocol~3 can tolerate only up to $f=5$
malicious agents. Recall that the models M2 and M3 share similar mobile behaviors,
where the difference is that in M2, agents in the cured rounds are aware.
However, Protocol~3 does not use this information and
thus works the same as the case of M3.
In both models, there may be $2f$ malicious values in the system
in each round. The success rates for all protocols go
to zero when $f_{\text{real}}>f$.

(iv)~Finally, we checked the performance under M3.
It is evident that except for Protocol~3, all protocols
fail to reach consensus even with one malicious agent.
As seen in Fig.~\ref{fig.g11},
Protocol~3 is capable when sufficient connectivity is available,
but cannot tolerate more malicious agents than the bound $f$.

We now turn our attention to the case $r=50$ with the smaller communication radius.
We have seen in Table~\ref{tab:add1}\,(a)
that Protocol~3 can deal with all models when $f_{\text{real}}\leq f$, but
this capability is realized by requiring a high level of connectivities.
It turns out that with $r=50$, the network is not connected enough
for Protocol~3, and it performs much worse than other protocols.
In fact, as shown in Table~\ref{tab:add1}\,(b),
Protocol~3 cannot reach consensus in any of the 10 topologies.
The conventional MSR, Protocols~1 and~2 have similar performance
as in the previous case with $r=70$.
Protocol~2A has some differences in M2 in that resilient consensus can be
guaranteed in all 10 topologies when $f_{\text{real}}\leq 2$.
Because of the mobile behavior in M2, the increase in $f_{\text{real}}$ can
lead to the increase in cured agents at each round.
Protocol~2A may have $2f_{\text{real}}$ cured agents in one round.
The cured agents do not send their values, which can reduce the connectivities.

\smallskip
We summarize the three simulation parts discussed in this section.
The malicious agents become more adversarial
according to the order in their models: The static, M1, M2, and M3.
Protocols designed for more adversarial models are capable
to deal with agents under less powerful models.
For example, all mobile protocols can handle the static
model, but the conventional MSR cannot reach consensus under any mobile models.
We also confirmed through these simulations that
protocols designed for more adversarial models require more
connectivities to guarantee resilient consensus.
These trade-offs are intuitive and can help the design
of network structures for resilient consensus.

%However, this in turn requires more connectivities
%for resilient consensus, and thus, when $r=50$, Protocol~3 performs
%much worse compared with other protocols.

\begin{table}[t]
    \centering
    \fontsize{8}{10}\selectfont
    \caption{The maximum of the actual number ${f_{\text{real}}}$ of
             malicious agents for achieving resilient consensus
             when $f=5$}
\label{tab:add1}
\vspace{-1mm}
{\small (a) With larger communication radius $r=70$}\\[1.5mm]
\begin{tabular}{c|cccc}
\hline
    %\toprule
%\diagbox [width=9em,trim=l] {Algorithm}{Model}
          & \multicolumn{4}{c}{Adversary model} \\ %\cline{2-5}
Algorithm &  Static       & M1 & M2 & M3 \\
\hline
Conventional MSR          &  $5$   &     $0$       &    $0$    &  $0$   \\
Protocol 1   &  $5$   &     $5$     &    $0$    &  $0$   \\
Protocol 2   &  $5$   &     $5$     &    $5$  &  $0$   \\
Protocol~2A   &  $5$   &     $5$     &    $5$  &  $0$   \\
Protocol 3   &  $10$  &     $10$    &    $5$  &  $5$ \\
    %\bottomrule
\hline
\end{tabular}%\vspace{0.5cm}
%\label{tab:add1}
%\end{table}
\vspace{6mm}

%\begin{table}[t]
%    \centering
%    \fontsize{8}{10}\selectfont
%    \caption{The maximum of the actual number ${f_{\text{real}}}$ of
%             malicious agents for achieving resilient consensus
%             when $f=5$ and $r=50$}
%\vspace{-2mm}
%    \caption{Resilient consensus
%           versus the actual number ${f_{\text{real}}}$
%           of malicious agents with $f=5$, $r=50$}
{\small (b) With smaller communication radius $r=50$}\\[1.5mm]
\begin{tabular}{c|cccc}
\hline
    %\toprule
    %\hline
%\diagbox [width=9em,trim=l] {Algorithm}{Model}&  Static       &         M1               &            M2        &         M3        \\
          & \multicolumn{4}{c}{Adversary model} \\ %\cline{2-5}
Algorithm &  Static       & M1 & M2 & M3 \\
\hline
Conventional MSR          &  $5$   &     $0$       &    $0$    &  $0$   \\
Protocol 1            &  $5$   &     $5$     &    $0$    &  $0$   \\
Protocol 2            &  $5$   &     $5$     &    $5$  &  $0$   \\
Protocol~2A            &  $5$   &     $5$     &    $2$  &  $0$   \\
Protocol 3            &     $-$             &     $-$                  &    $-$               &  $-$              \\
    %\bottomrule
\hline
\end{tabular}%\vspace{0cm}
%    \label{tab:add2}
\end{table}

\section{Conclusion}
\label{Section 8}

In this paper, we have considered the multi-agent consensus
problem in the presence of mobile misbehaving agents and have developed
resilient protocols to mitigate their influence on the regular agents.
Specifically, under three classes of mobile malicious agents, four
protocols have been proposed. For the protocols to achieve resilient consensus,
we have characterized the conditions on the necessary graph structures
through theoretical analyses under networks in both complete and non-complete
graph forms. We have observed that these conditions reflect
the different levels of adversarial capabilities that the three classes of
mobile malicious agents possess.
By means of numerical simulations, we have further studied
the performance of the proposed resilient consensus protocols
for random networks of 100 nodes where the theoretical conditions
may not hold.

In future research, we will focus on formulating more detailed models
for mobile adversary behaviors. We would also like to extend our approach
to other multi-agent tasks where the adversary's mobile capabilities
may create complexity in the responses and actions of the regular agents
for protecting the overall system. Furthermore, asynchronous update behaviors
as well as time delays in communication should be taken into account.

%\appendices
%\section{Proof of the First Zonklar Equation}
%Appendix one text goes here.

% you can choose not to have a title for an appendix
% if you want by leaving the argument blank
%\section{}
%Appendix two text goes here.

% use section* for acknowledgment
%\ifCLASSOPTIONcompsoc
  % The Computer Society usually uses the plural form
%  \section*{Acknowledgments}
%\else
  % regular IEEE prefers the singular form
%  \section*{Acknowledgment}
%\fi

%The authors would like to thank...

\ifCLASSOPTIONcaptionsoff
  \newpage
\fi

\begin{IEEEbiography}[%
%{\includegraphics[width=1in,height=1.25in,clip,keepaspectratio]{wang.jpg}}
]{Yuan~Wang}
Yuan Wang received the M.Sc.\ degree in engineering from
Huazhong University of Science and Technology, Wuhan,
China in 2016, and the Ph.D.\ degree in Artificial
Intelligence from Tokyo Institute of Technology, Yokohama,
Japan in 2019. He is currently a researcher in the
Department of Computer Science, Tokyo Institute of Technology,
Yokohama, Japan.

His main research interests are cyber-physical
systems, event-based coordination, security in multi-agent systems,
and model predictive control methods.
\end{IEEEbiography}

\begin{IEEEbiography}[%
%{\includegraphics[width=1in,height=1.25in,clip,keepaspectratio]{ishii.jpg}}
]{Hideaki Ishii} (M'02-SM'12) received the M.Eng.\ degree in
applied systems science from Kyoto University, Kyoto, Japan,
in 1998, and the Ph.D.\ degree in electrical and computer
engineering from the University of Toronto, Toronto, ON,
Canada, in 2002.
He was a Postdoctoral Research Associate with
the Coordinated Science Laboratory at the University
of Illinois at Urbana-Champaign, Urbana, IL, USA,
from 2001 to 2004, and a Research Associate with the
Department of Information Physics and Computing,
The University of Tokyo, Tokyo, Japan, from 2004 to 2007.
Currently, he is an Associate Professor in the Department
of Computer Science, Tokyo Institute of Technology, Yokohama, Japan.
His research interests are in networked control systems,
multiagent systems, cyber security of power systems,
and distributed and probabilistic algorithms.

Dr.~Ishii has served as an Associate Editor for
the \emph{IEEE Control Systems Letters} and
the \emph{Mathematics of Control, Signals, and Systems}
and previously for \emph{Automatica}, the \emph{IEEE Transactions
on Automatic Control}, and the \emph{IEEE Transactions on Control
of Network Systems}.
He is the Chair of the IFAC Coordinating Committee on Systems and Signals
since 2017.
He received the IEEE Control Systems Magazine Outstanding
Paper Award in 2015.
\end{IEEEbiography}

\begin{IEEEbiography}[%
]{Fran\c{c}ois~Bonnet}
Fran\c{c}ois Bonnet is a Specially Appointed
Associated Professor at Tokyo Tech since 2018.

He obtained his M.S.\ from the ENS Cachan at Rennes,
France in 2006 and his Ph.D.\ from the University of
Rennes~1 in 2010. He worked at JAIST as a JSPS postdoctoral
fellow until 2012 and then as an Assistant Professor until 2017.
Then he spent one year as a Specially Appointed Assistant
Professor at Osaka University.

His research interests include theoretical distributed
computing, discrete algorithms, and (combinatorial) game
theory.
\end{IEEEbiography}

\begin{IEEEbiography}[
%{\includegraphics[width=1in,height=1.25in,clip,keepaspectratio]{defago.jpg}}
]{Xavier~D\'{e}fago}
Xavier D\'{e}fago is a full professor at Tokyo
Institute of Technology since 2016.

He obtained master (1995) and PhD (2000) in computer science from the Swiss Federal Institute of Technology in Lausanne (EPFL) in Switzerland. Before his current position at Tokyo Tech, he was a faculty member at the Japan Advanced Institute of Science and Technology (JAIST).
Meanwhile, he has also been a PRESTO researcher for the Japan Science and Technology Agency (JST), and an invited researcher for CNRS (France) at Sorbonne University and at INRIA Sophia Antipolis.

He is a member of the IFIP working group 10.4 on dependable computing and fault-tolerance. He served as program chair of IEEE SRDS in 2014 and IEEE ICDCS in 2012 and as general chair of SSS 2018.

Xavier has been working on various aspects of dependable computing such as distributed agreement, state machine replication, failure detection, and fault-tolerant group communication in general.
His interest include also robotics, embedded systems, and programming languages.
\end{IEEEbiography}

\end{document}